\documentclass[useAMS, usenatbib, onecolumn]{mn2e}
\usepackage{epsfig}
\usepackage{amsmath}
\def\be{\begin{equation}}
\def\ee{\end{equation}}
\def\ba{\begin{eqnarray}}
\def\ea{\end{eqnarray}}
\def\go{\mathrel{\raise.3ex\hbox{$>$}\mkern-14mu
             \lower0.6ex\hbox{$\sim$}}}
\def\lo{\mathrel{\raise.3ex\hbox{$<$}\mkern-14mu
             \lower0.6ex\hbox{$\sim$}}}

\def\bu{{\bf u}}

\def\cR{{\cal R}}
\def\d{\partial}

\def\tomega{\tilde\omega}
\def\tomega{{\tilde{\omega}}}

\begin{document}
\title[Modes in Black Hole Accretion Discs]
{Corotational Instability of Inertial-Acoustic Modes in Black Hole
Accretion Discs and Quasi-Periodic Oscillations}
\author[D. Lai and D. Tsang]
{Dong Lai$^{1}$\thanks{Email:
dong@astro.cornell.edu; dtsang@astro.cornell.edu} 
and David Tsang$^{1,2}$\footnotemark[1] \\ 
$^1$Department of Astronomy, Cornell University, Ithaca, NY 14853, USA \\
$^2$Department of Physics,
Cornell University, Ithaca, NY 14853, USA \\}

\label{firstpage}
\maketitle

\begin{abstract}
We study the global stability of non-axisymmetric p-modes
(also called inertial-acoustic modes) trapped in the inner-most regions
of accretion discs around black holes.  We show that the lowest-order
(highest-frequency) p-modes, with frequencies $\omega=(0.5-0.7)
m\Omega_{\rm ISCO}$ (where $m=1,2,3,\cdots$ is the azimuthal wave
number, $\Omega_{\rm ISCO}$ is the disc rotation frequency at the
Inner-most Stable Circular Orbit, ISCO), can be overstable due to 
general relativistic effects, according to which the radial epicyclic
frequency $\kappa$ is a non-monotonic function of radius near the black
hole. The mode is trapped inside the corotation resonance radius $r_c$ 
(where the wave pattern rotation speed $\omega/m$
equals the disc rotation rate $\Omega$) and carries a negative energy. 
The mode growth arises primarily from wave absorption at the
corotation resonance, and the sign of the wave
absorption depends on the gradient of the disc vortensity,
$\zeta=\kappa^2/(2\Omega\Sigma)$ (where $\Sigma$ is the surface
density). When the mode frequency $\omega$ is
sufficiently high, such that $d\zeta/r>0$ at $r_c$,
positive wave energy is absorbed at the corotation, leading to the
growth of mode amplitude. The mode growth is further enhanced by
wave transmission beyond the corotation barrier.  We also study how
the rapid radial inflow at the inner edge of the disc affects the mode
trapping and growth. Our analysis of the behavior of the fluid perturbations
in the transonic flow near the ISCO indicates that, while
the inflow tends to damp the mode, the damping effect is sufficiently
small under some conditions (e.g., when the disc density decreases
rapidly with decreasing radius at the sonic point) so that net mode growth can 
still be achieved. We further clarify the role of the Rossby wave
instability and show that it does not operate for black hole accretion
discs with smooth-varying vortensity profiles.
Overstable non-axisymmetric p-modes driven by the
corotational instability provide a plausible explanation for the
high-frequency ($\go 100$~Hz) quasi-periodic oscillations (HFQPOs) 
observed from a number of black-hole X-ray binaries in the very high 
state. The absence of HFQPOs in the soft (thermal) state may result from 
mode damping due to the radial infall at the ISCO.
\end{abstract}

\section{Introduction}

\subsection{Models of Quasi-Periodic X-Ray Oscillations: A Brief Review}

Rapid X-ray variabilities from Galactic compact binary systems have
been studied for decades (e.g. van del Klis 2006).  In recent years,
our knowledge of quasi-periodic oscillations (QPOs) in black-hole
X-ray binaries has greatly improved (see Remillard \& McClintock 2006
for a review), thanks in large part to NASA's {\it Rossi X-ray Timing
Explorer} (Swank 1999). The low-frequency QPOs (about 0.1-50~Hz) are
common, observable when the systems are in the hard state and the
steep power-law state (also called ``very high state''; see Done,
Gierlinski \& Kubota 2007), and they typically have high amplitudes
and high coherence ($Q>10$), and can vary in frequency on short
timescales (minutes).  However, it is the weaker, transient High-Frequency QPOs
(HFQPOs, 40--450~Hz) that have attracted more attention, since their
frequencies do not vary significantly in response to sizable (factors
of 3-4) luminosity changes and are comparable to the orbital
frequencies at the Innermost Stable Circular Orbit (ISCO) of black
holes with mass $M\sim 10 M_{\odot}$. As such, HFQPOs potentially
provide a probe to study the effects of strong gravity. HFQPOs are
usually observed in the very high state of the X-ray binaries, 
and have low amplitudes
($0.5-2\%$ rms at 2-60~keV) and low coherence ($Q\sim 2-10$). Out of
the seven black-hole binaries from which HFQPOs have been reported, four
show pairs of QPOs (first discovered in GRO J1655-40; Strohmayer 2001)
with frequency ratios close to $2:3$ (300 and
450~Hz in GRO J1655-40, 184 and 276~Hz in XTE J1550-564, 113 and
168~Hz in GRS 1915+105, 165 and 240~Hz in H1743-322; note that GRS
1915+105 also has a second pair of QPOs with $f=41$ and $67$~Hz).

It is worth noting that QPO (with period of $\sim 1$~hour) in X-ray emission
has recently been detected in the active galaxy RE J1034+396 
(Gierlinski et al.~2008). This could be the ``supermassive'' analog of the
HFQPOs detected in black-hole X-ray binaries.

Despite the observational progress, the origin of the HFQPOs remain 
unclear. A number of possibilities/models have been suggested or
studied to various degrees of sophistication. We comment on some of these
below:

-- Stella, Vietri \& Morsink (1999) and others (see Schnittman \&
Bertschinger 2004, Schnittman 2005) suggested that orbiting hot spots
(blobs) in the disc oscillating with epicyclic frequencies may provide
variability in the X-ray emission.  However the radial positions of
such blobs are free parameters, which must be tuned to match the
observed QPO frequencies, and it is also not clear that the blobs can
survive the differential rotation of the disc.

-- Abramowicz \& Kluzniak (2001) suggested that HFQPOs involve certain
nonlinear resonant phenomenon in the disc (e.g., coupling between the
radial and vertical epicyclic oscillations of the disc fluid element;
Kluzniak \& Abramowicz 2002).  This was motivated by the observed
stability of the QPO frequencies and the commensurate frequency ratio.
However, so far analysis has been done based only on toy models involving coupled harmonic oscillators (e.g. Rebusco 2004; Horak \& Karas 2006) and no fluid dynamical model producing these resonances
has been developed (see Abramowicz et al 2007 and Rebusco 2008 for
recent reviews). Petri (2008) considered the resonant oscillation of a
test mass in the presence of a spiral density wave, but the origin of
the wave is unclear.

-- Acoustic oscillation modes in pressure-supported accretion tori
have been suggested as a possible source of the observed QPOs
(Rezzolla et al. 2003; Lee, Abramowicz \& Kluziniak 2004; see also
Blaes, Arras \& Fragile 2006; Schnittman \& Rezzolla 2006, Blaes 
et al 2007, Sramkova et al 2007).  In this
model, the commensurate mode frequencies arise from matching the
radial wavelength to the size of the torus. Note that the QPO
frequencies are determined mainly by the radial boundaries of the
torus, which must be tuned to match the observed QPO frequencies. It
is also not clear that the accretion flow in the very high state (in
which HFQPOs are observed) is well represented by such a torus (e.g. Done et al 2007).

-- Li \& Narayan (2004) considered the dynamics of the
interface between the accretion
disc and the magnetosphere of a central compact object (see also
Lovelace \& Romanova 2007). 
The interface is generally Rayleigh-Tayor unstable and may also be
Kelvin-Helmholtz unstable. While such an interface is clearly relevant
to accreting magnetic neutron stars, Li \& Narayan suggested that it
may also be relevant to accreting black holes and that the strongly
unstable interface modes may give rise to QPOs with commensurate
frequencies.

-- Perhaps the theoretically most appealing is the relativistic
diskoseismic oscillation model, according to which general
relativistic (GR) effects produce trapped oscillation modes in the
inner region of the disc (Kato \& Fukue 1980; Okazaki et al.~1987;
Nowak \& Wagoner 1991; see Wagoner 1999; Kato 2001 for reviews; see
also Tassev \& Bertschinger 2007 for the kinematic description of some
of these wave modes).  A large majority of previous studies have
focused on disc g-modes (also called inertial modes or
inertial-gravity modes, whose wavefunctions -- such as the pressure
perturbation, contain nodes in the vertical direction), because the
trapping of the g-mode does not require a reflective inner/outer disc
boundary.  Kato (2003a) and Li, Goodman \& Narayan (2003) showed that
the g-mode that contains a corotation resonance (where the wave
patten frequency equals the rotation rate of the background flow) in
the wave zone is heavily damped. Thus the only nonaxisymmetric ($m\neq
0$) g-modes of interest are those trapped around the maximum of
$\Omega+\kappa/m$ (where $\Omega$ is the rotational frequency,
$\kappa$ is the radial epicyclic frequency and $m$ is the azimuthal
mode number; see Fig.~1 below). Unfortunately, the frequencies of such
modes, $\omega\simeq m\Omega_{\rm ISCO}$, are too high (by a
factor of 2-3) compared to the observed values, given the measured
mass and the estimated spin parameter of the black hole (Silbergleit
\& Wagoner 2007; see also Tassev \& Bertschinger 2007).  Axisymmetric
g-modes ($m=0$) may still be viable in the respect, and recent studies
showed that they can be resonantly excited by global disc deformations
through nonlinear effects (Kato 2003a,2008; Ferreira \& Ogilvie 2008).
Numerical simulations (Arras, Blaes \& Turner 2006; Reynolds \& Miller
2008), however, indicated that while axisymmetric g-mode oscillations
are present in the hydrodynamic disc with no magnetic field, they
disappear in the magnetic disc where MHD turbulence develops. Also, Fu
\& Lai (2008) carried out an analytic study of the effect of magnetic
fields on diskoseismic modes and showed that even a weak (sub-thermal)
magnetic field can ``destroy'' the self-trapping zone of disc g-modes,
and this may (at least partly) explain the disappearance of the
g-modes in the MHD simulations.

-- Tagger and collaborators (Tagger \& Pellat 1999; Varniere \& Tagger
2002; Tagger \& Varniere 2006; see Tagger 2006 for a review) developed
the theory of accretion-ejection instability for discs threaded by
strong (of order or stronger than equipartition), large-scale poloidal
magnetic fields.  They showed that such magnetic field provides a
strong coupling between spiral density waves and Rossby waves at the
corotation, leading to the growth of the waves and energy ejection to
disc corona. Tagger \& Varniere (2006) suggested that normal modes
trapped in the inner region of the disc become strongly unstable by a
combination of accretion-ejection instability and an MHD form of the
Rossby wave instability (see Lovelace et al.~1999; Li et al.~2000; see section 6
below). The Tagger model has the appealing feature that the instability leads
to energy ejection into the disc corona, and thus explains why HFQPOs
manifest mainly as the variations of the nonthermal (power-law)
radiation from the systems.


\subsection{This Paper}

In this paper we study the global corotational instability of
nonaxisymmetric p-modes (also called inertial-acoustic modes) trapped
in the inner-most region of the accretion disc around a black
hole. The p-modes do not have vertical structure (i.e., the
wavefunctions have no node in the vertical direction). We focus on
these modes because their basic wave properties (e.g. propagation
diagram) are not affected qualitatively by disc magnetic fields (Fu \&
Lai 2008) and they are probably robust under hydromagnetic effects
and disc turbulence (see Reynolds \& Miller 2008).

The corotational instability of p-modes studied in this paper relies
on the well-known GR effect of test-mass orbit around a black hole:
Near the black hole, the radial epicyclic frequency $\kappa$ reaches a
maximum (at $r=8GM/c^2$ for a Schwarzschild black hole) and goes to
zero at the ISCO ($r_{\rm ISCO}=6GM/c^2$).  This causes non-monotonic
behavior in the fluid vortensity, $\zeta=\kappa^2/(2\Sigma\Omega)$
(assuming the surface density $\Sigma$ is relatively smooth), such
that $d\zeta/dr>0$ for $r<r_{\rm peak}$ (where $r_{\rm peak}$ is the
radius where $\zeta$ peaks) and $d\zeta/dr<0$ for $r>r_{\rm peak}$.
The vortensity gradient $d\zeta/dr$ plays an important role in wave
absorption at the corotation resonance (Tsang \& Lai 2008a; see
also Goldreich \& Tremaine 1979 for corotational wave absorption due to
external forcing). We show that the p-modes with frequencies such that
the corotation radii lie inside the vortensity peak can grow in
amplitude due to absorption at the corotation resonance, and that the
overstability can be achieved for several modes with frequencies
closely commensurate with the azimuthal wavenumber $m$.  Tagger \&
Varniere (2006) have studied similar modes in discs threaded by strong
magnetic fields, but in our analysis magnetic fields play no role.

The trapping of the p-modes requires the existence of a (partially)
reflecting boundary at the disc inner edge, close to the ISCO. One may
suspect that the rapid radial inflow at the ISCO will diminish any
potential instabilities in the inner accretion disc (see Blaes 1987
for the case of thick accreting tori). Our analysis of the
wave perturbations in the transonic accretion flow (see section 5)
suggests that waves are partially reflected at the sonic point, and
global overstable p-modes may still be produced under certain 
conditions (e.g., when the surface density of the flow
varies on sufficiently small length scale around the sonic point).
Even better mode trapping (and therefore larger mode growth)
may be achieved when the system is an accretion
state such that the inner disc edge does not behave as a zero-torque
boundary (see section 7 for discussion and references).

Our paper is organized as follows. After summarizing the basic
fluid equations for our problem in section 2, we give a physical
discussion of the origin of the corotational instability of disc 
p-modes in section 3. We present in section 4
our calculations of the growing p-modes with simple reflective
inner disc boundary conditions. Section 5 contains our analysis of
the effect of the transonic radial inflow at the ISCO 
on the p-mode growth rate. In section 6, we discuss the role of the Rossby 
wave instability and show that it is not effective in typical 
accretion discs under consideration.
In section 7 we discuss the application
of our results to HFQPOs in black hole X-ray binaries.

\section{Setup and Basic equations}

We consider a geometrically thin disc and adopt cylindrical
coordinate system $(r,\phi,z)$. The flow is assumed to be barotropic,
so that the vertically integrated pressure, $P=\int p~dz$, depends
only on the surface density, $\Sigma=\int \rho~dz$.  We use
the pseudo-Newtonian potential of Paczynski \& Wiita (1980) 
\be
\Phi = -{GM\over r-r_S},
\ee
with $r_S = 2GM/c^2$ the Schwarzschild radius.
The free-particle (Keplerian) orbital and radial epicyclic (angular)
frequencies are
\be
\Omega_K = \left(\frac{1}{r}\frac{d\Phi}{dr}\right)^{1/2} =
\sqrt{\frac{GM}{r}}\frac{1}{r-r_S}~, \qquad \kappa =
\left[\frac{2\Omega_K}{r} \frac{d}{dr}(r^2\Omega_K)\right]^{1/2} = \Omega_K
\sqrt{\frac{r-3r_S}{r-r_S}}.
\label{eq:OmegaK}
\ee
The function $\kappa$ peaks at $r=(2+\sqrt{3})r_S$ and declines to zero at 
$r_{\rm ISCO}=3r_S$ (while for a Schwarzschild black hole in GR, 
$\kappa$ peaks at $r=4r_S$). 
The unperturbed flow has velocity $\bu_0=(u_r,r\Omega,0)$.
Since pressure is negligible for thin discs, we have
$\Omega\simeq \Omega_K$.

Neglecting the self-gravity of the disc we have the linear
perturbation equations:
\ba
&& {\partial\over\partial t}\delta\Sigma+\nabla\cdot(\Sigma\,\delta \bu
+\bu_0\,\delta\Sigma)=0,\label{eq:rho}\\
&&{\partial \over\partial t}\delta\bu+(\bu_0\cdot\nabla)\delta\bu
+(\delta\bu\cdot\nabla)\bu_0
=-\nabla\delta h, \label{eq:u}
\ea
where $\delta\Sigma,~\delta \bu$ and $\delta h=\delta P/\Sigma$
are the (Eulerian) perturbations of surface density, velocity and enthalpy,
respectively. For barotropic flow, $\delta h$ and $\delta\Sigma$ are 
related by
\be
\delta h=c_s^2{\delta\Sigma\over\Sigma},
\ee
where $c_s$ is the sound speed, with $c_s^2=dP/d\Sigma$.
We assume all perturbed quantities to be of the form
$e^{im\phi - i\omega t}$, 
where $m$ is a positive integer, and $\omega$ is the 
wave (angular) frequency.
The perturbation equations then become
\ba
&& -i\tomega {\Sigma\over c_s^2}\delta h + \frac{1}{r} \frac{\d}{\d r}
(\Sigma r \delta u_r) + \frac{im}{r}\Sigma \delta u_\phi 
+{1\over r}{\partial\over\partial r}(ru_r\delta\Sigma)
= 0,\label{eq:sig}\\
&& -i\tomega \delta u_r - 2\Omega \delta u_\phi 
+{\partial\over\partial r}(u_r\delta u_r)= -{\partial\over\partial r}
\delta h,\label{eq:ur}\\
&& -i\tomega \delta u_\phi + \frac{\kappa^2}{2\Omega}\delta u_r 
+{u_r\over r}{\partial\over\partial r}(r\delta u_\phi)
=-\frac{im}{r} \delta h,\label{eq:uphi}
\ea
where 
\be
\tomega = \omega - m \Omega,
\ee
is the wave frequency in the frame corotating with the unperturbed fluid.

Except very near the inner edge of the disc, $r_{\rm in}\simeq r_{\rm ISCO}$,
the unperturbed radial velocity is small, $|u_r|\ll r\Omega$.
In our calculations of the disc modes, we will neglect $u_r$ and
set the last terms on the left-hand sides of 
equations (\ref{eq:sig})-(\ref{eq:uphi}) to zero (However,
$u_r$ plays an important role in determining the inner boundary 
condition of the fluid perturbations at $r_{\rm in}$; see section 5).
Eliminating the velocity perturbations in favor of the enthalpy,
we obtain our master equation
\be
\left[\frac{d^2}{dr^2} -
\frac{d}{dr}\left(\ln\frac{D}{r\Sigma}\right)\frac{d}{dr} -
\frac{2m\Omega}{r\tomega}\left(\frac{d}{dr}\ln\frac{\Omega\Sigma}{D}\right)
- \frac{m^2}{r^2} - \frac{D}{c_s^2}\right]\delta h = 0,
\label{perturbeq1}
\ee
where 
\be
D = \kappa^2 - \tomega^2.
\ee
For concreteness we assume the surface density to have a the power-law form 
\be
\Sigma \propto r^{-p},
\ee
where $p$ is the density index.

The above equations adequately describe disc p-modes (also called
inertial-acoustic modes), which do not have vertical structure.
Other disc modes (g-modes and c-modes) involve the vertical
degree of freedom [see Kato 2001 for a review; also see Fig.~1 of
Fu \& Lai (2008) for a quick summary],
and their stability properties are studied by
Kato (2003a), Li et al.~(2003) and Tsang \& Lai (2008b).

To determine the global modes of the disc, appropriate 
boundary conditions must be specified. These are discussed in 
sections 4 and 5.

\section{P-modes and Their Growth Due
to Corotation Resonance: A Physical Discussion}

A WKB analysis of the wave equation (\ref{perturbeq1})
yields the dispersion 
relation for the local plane wave 
$\delta h\propto \exp\left[i\int^r\!k(s) ds\right]$:
\be
k^2+{D\over c_s^2}+
\frac{2m\Omega}{r\tomega}\left(\frac{d}{dr}\ln\frac{\Omega\Sigma}{D}\right)
\simeq 0.
\label{eq:disper}\ee
Far from the singularity ($\tomega=0$) at the corotation radius $r_c$, 
this reduces to the 
well-known dispersion relation of spiral density wave with no self-gravity
(e.g., Shu 1992), $k^2\simeq -D/c_s^2$, or 
\be
\tomega^2 = \kappa^2 + k^2 c_s^2~.
\ee
Density waves (p-modes) can propagate inside the inner Lindblad resonance
radius $r_{\rm IL}$ (defined by $\tomega = -\kappa$), and outside the
outer Lindblad resonance radius $r_{\rm OL}$ (defined by $\tomega =
\kappa$), i.e., in the region where
$\omega/m<\Omega-\kappa/m$ and 
$\omega/m > \Omega+\kappa/m$, respectively (see Fig.~1). 
Between $r_{\rm IL}$ and $r_{\rm OL}$, waves are 
evanescent except that a very narrow Rossby wave zone exists 
around the corotation radius. Indeed, 
in the vicinity of $\tomega=0$, equation (\ref{eq:disper}) reduces to 
\be
\tomega\simeq {2m\Omega\over r (k^2+\kappa^2/c_s^2)}
\left({d\ln\zeta\over dr}\right)_{r_c},
\ee
where 
\be
\zeta={\kappa^2\over 2\Omega\Sigma}
\ee
is the vortensity of the (unperturbed) flow.
For $(d\zeta/dr)_{r_c}>0$, the Rossby wave zone lies between
$r_c$ and $r_c+\Delta r_R$, where $\Delta r_R=(2c_s/\kappa)|\nu|$ and
$\int_{r_c}^{r_c+\Delta r_R}k\,dr=\pi\nu$, with the number of wavelengths in the
Rossby zone given by (Tsang \& Lai 2008a)
\be
\nu = \left(\frac{c_s}{q\kappa}\frac{d\ln \zeta}{dr} \right)_{r_c}
={c_s\over q\kappa}\left[{d\over dr}\ln\left({\kappa^2\over\Omega}\right)
+{p\over r}\right]_{r_c},
\label{eq:nu}\ee
where $q \equiv -\left(d\ln \Omega/d \ln r\right)_{\rm r_c}$, and the second
equality assumes $\Sigma\propto r^{-p}$. 
For $(d\zeta/dr)_{r_c}<0$, the Rossby wave zone lies inside $r_c$, between
$r_c-|\Delta r_R|$ and $r_c$ (see Fig.~1). Note that since $\nu\sim c_s/(\kappa r)\sim
H/r\ll 1$, no standing Rossby wave can exist in the Rossby zone (see also section 6).

\begin{figure}
\centering
\includegraphics[width=13.5cm]{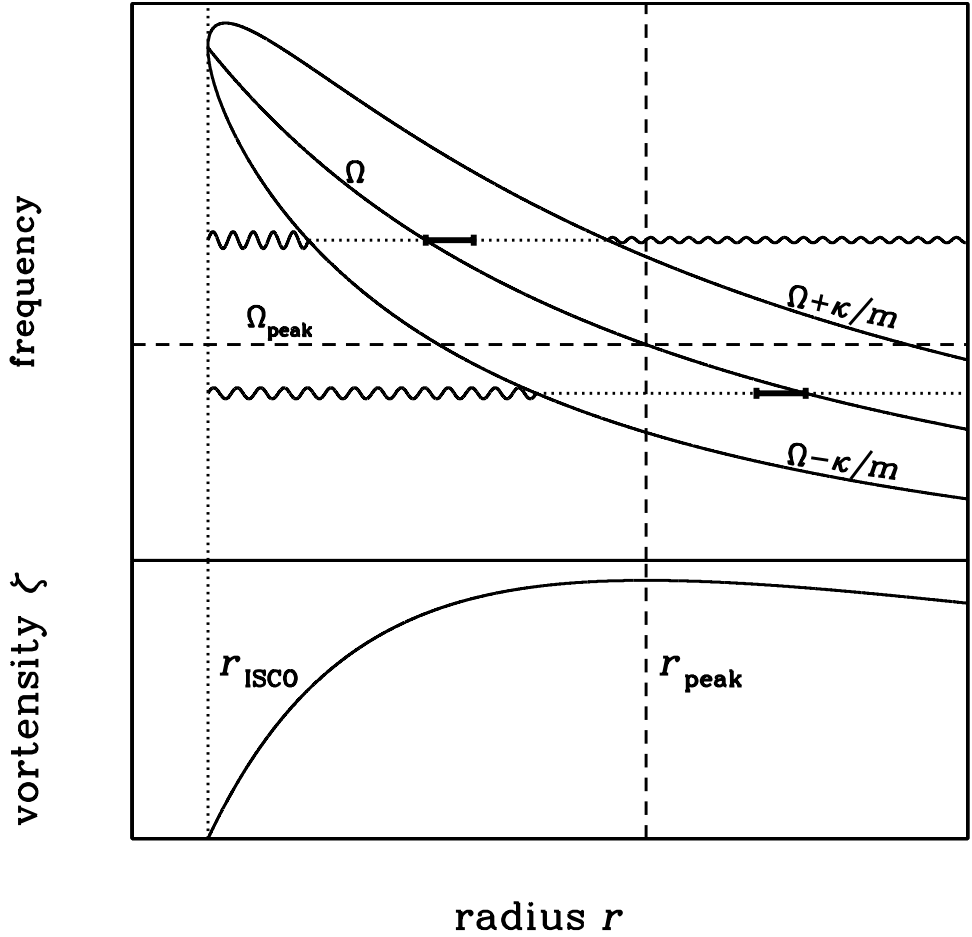}
\caption{Wave propagation diagram for non-axisymmetric p-modes in thin
accretion discs around black holes.  In the upper panel, the three
solid curves depict the disc rotation profile $\Omega(r)$ and
$\Omega\pm\kappa/m$ (where $\kappa$ is the radia epicyclic frequency);
note that the three curves join each other at the disc inner radius
$r_{\rm in}=r_{\rm ISCO}$ (the inner-most circular orbit) since
$\kappa(r_{\rm ISCO})=0$.  The wavy lines (of height $\omega/m$)
indicate the propagation zones for inertial-acoustic waves. Disc
p-modes are trapped between $r_{\rm in}$ and the inner Lindblad
resonance radius (where $\omega/m = \Omega - \kappa/m$), but can
tunnel through the corotation barrier. The lower panel depicts the
disc vortensity profile, $\zeta=\kappa^2/(2\Omega\Sigma)$, which has a
maximum at the radius $r_{\rm peak}$ (as long as the surface density
$\Sigma$ does not vary too strongly with $r$).  P-modes with
$\omega/m>\Omega_{\rm peak}=\Omega(r_{\rm peak})$ (the upper wavy
line) are overstable due to wave absorption at the corotation
resonance radius $r_c$ (where $\omega/m=\Omega$) since
$(d\zeta/dr)_{r_c}>0$. P-modes with $\omega/m<\Omega_{\rm peak}$ (the
lower wavy line) tend to be damped by wave absorption at $r_c$ since
$(d\zeta/dr)_{r_c}<0$. Note that a narrow Rossby wave zone (labeled by
thick horizontal bars) exists just outside or inside the corotation
radius --- the location of this Rossby zone determines the sign of
the corotational wave absorption.  }
\label{fig1}
\end{figure}

Assuming that there exists a reflecting boundary at the inner 
disc radius $r_{\rm in}\simeq r_{\rm ISCO}$ (see sections 4.3 and 5),
normal modes can be produced, with the waves partially trapped between 
$r_{\rm in}$ and 
$r_{\rm IL}$ -- these are the p-modes that we will focus on in this paper.
The mode eigen-frequency $\omega=\omega_r+i\omega_i$ is generally complex,
with the real part $\omega_r$ determined approximately by the Sommerfeld 
``quantization'' condition
\be
\int_{r_{\rm in}}^{r_{\rm IL}} {1 \over c_s} \sqrt{\tomega_r^2
- \kappa^2} dr = n\pi +\varphi, 
\ee 
where $\tomega_r=\omega_r-m\Omega$, 
$n$ is an integer and $\varphi$ (of order unity) is a phase factor
depending on the details of the (inner and outer) boundary conditions.
The overstability of the p-mode is directly related to the 
reflectivity of the corotation barrier between $r_{\rm IL}$ and 
$r_{\rm OL}$. In the WKB approximation, the imaginary part of the 
mode frequency is given by (Tsang \& Lai 2008a; see also Narayan et al.~1987,
who considered shearing-sheet model)
\be
\omega_i = \left(\frac{|{\cal R}| -1}{|{\cal R}|+1}\right)
\left(\int_{r_{\rm in\ }}^{r_{\rm IL}}
{|\tomega_r|\over c_s\sqrt{\tomega_r^2 - \kappa^2}} dr\right)^{-1},  
\label{eq:omegai}\ee
where ${\cal R}$ is the reflectivity (see below).
Thus the mode becomes overstable ($\omega_i > 0$) for $|{\cal R}| > 1$
(termed ``super-reflection'')  and stable
($\omega_i < 0$) for $|{\cal R}| < 1$.

Super-reflection in fluid discs arises because
the waves inside the corotation radius and those outside carry 
energy or angular momentum of different signs: 
Since the wave inside $r_c$ has pattern speed $\omega_r/m$ less than the 
fluid rotation rate $\Omega(r)$, it carries negative energy; outside
$r_c$, we have $\omega_r/m>\Omega(r)$, the wave carries positive 
energy. Consider an incident wave $\delta h\propto \exp(-i\int^r k\,dr)$, 
carrying energy of the amount $(-1)$, propagating from small radii
toward the corotation barrier
\footnote{Note that since the group velocity 
of the wave has opposite sign as the phase velocity for $r<r_{\rm IL}$,
the wave of the form $\exp(-i\int^rk\,dr)$ (with $k>0$)
is outward propagating.}. The wave reflected at $r_{\rm IL}$
takes the form $\delta h\propto {\cal R}\exp(i\int^r k\,dr)$, and the 
transmitted wave in the region $r>r_{\rm OL}$ is
$\delta h\propto {\cal T}\exp(i\int^r k\,dr)$. 
Because of the corotation singularity, the wave energy can also be
transferred to the background flow and dissipated at the corotation 
radius.   Energy conservation then gives
$-1=(-1)|{\cal R}|^2+|{\cal T}|^2+{\cal D}_c$, or
\be
|{\cal R}|^2=1+|{\cal T}|^2+{\cal D}_c, \label{eq:Econs}
\ee
where ${\cal D}_c$ is the wave energy dissipated at the corotation. 

Tsang \& Lai (2008a) derived the analytical expressions (in the WKB
approximation) for ${\cal T}$, ${\cal R}$ and ${\cal D}_c$. 
Two effects determine the reflectivity.
(i) The transmitted wave (corresponding to the $|{\cal T}|^2$ term)
always carries away positive energy and thus increases 
$|{\cal R}|^2$.
(ii) Wave absorption at the corotation can have both signs, depending on 
$\nu$: For $\nu>0$, the Rossby wave zone lies outside $r_c$, positive wave
energy is dissipated and we have ${\cal D}_c>0$; for $\nu<0$, the Rossby zone
lies inside $r_c$ and we have ${\cal D}_c<0$ (see Fig.~1). 
Tsang \& Lai (2008a) showed explicitly that under most conditions,
$|{\cal D}_c|\gg |{\cal T}|^2$ (except when $\nu\simeq 0$, for which
${\cal D}_c\simeq 0$). In the limit of $|\nu|\ll 1$, we have
\be
|{\cal T}|^2 \simeq \exp(-2\Theta_{\rm II}),\qquad
{\cal D}_c\simeq 2\pi\nu\exp(-2\Theta_{\rm IIa}),
\ee
where
\be
\Theta_{\rm II}\equiv \int_{r_{\rm IL}}^{r_{\rm OL}}
\frac{\sqrt{\kappa^2 - \tomega_r^2}}{c_s}, \qquad \Theta_{\rm IIa}\equiv
\int_{r_{\rm IL}}^{r_{c}} \frac{\sqrt{\kappa^2 - \tomega_r^2}}{c_s}
\ee
[these expressions are valid for $\Theta_{\rm II},\Theta_{\rm IIa}\gg 1$; see 
Tsang \& Lai (2008a) for more general expressions]. Thus super-reflectivity
($|{\cal R}|^2>1$) and growing modes ($\omega_i>0$) are achieved when 
\be
\nu>\nu_{\rm crit}=-{1\over 2\pi}\exp(-2\Theta_{\rm IIb}),\qquad
{\rm with}\quad\Theta_{\rm IIb}=\Theta_{\rm II}-\Theta_{\rm IIa}
=\int_{r_c}^{r_{\rm OL}}
\frac{\sqrt{\kappa^2 - \tomega_r^2}}{c_s}.
\label{eq:nucrit}\ee
Note that typically $|\nu_{\rm crit}|\ll 1$; if $|{\cal T}|^2$ is neglected
compared to ${\cal D}_c$, then $\nu_{\rm crit}=0$.

As mentioned before, since $\kappa$ is non-monotonic near the black
hole, the vortensity $\zeta$ is also non-monotonic, attaining a peak
value at $r=r_{\rm peak}$ before dropping to zero at the
ISCO. Therefore, p-modes with frequencies such that the corotation
radius $r_c$ lies inside $r_{\rm peak}$ are expected to be overstable
by the corotational instability discussed above. In other words, when
$\omega_r/m> \Omega_{\rm peak}\equiv \Omega(r_{\rm peak})$, the
corotation resonance acts to grow the mode.  Note that $\Omega_{\rm
peak}$ depends on the surface density profile as well as the spacetime
curvature around the black hole (see Fig.~2).  On the other hand, when
$\omega_r/m< \Omega_{\rm peak}$ ($\nu<0$), the corotational wave
absorption acts to damp the mode.  However, when $\omega_r/m$ is only
slightly smaller than $\Omega_{\rm peak}$ ($\nu_{\rm crit}<\nu<0$)
mode growth can still be obtained due to wave leakage beyond the outer
Lindblad resonance, though the growth rate will be small (see section
4.4 for examples).

\begin{figure}
\centering
\includegraphics[width=13cm]{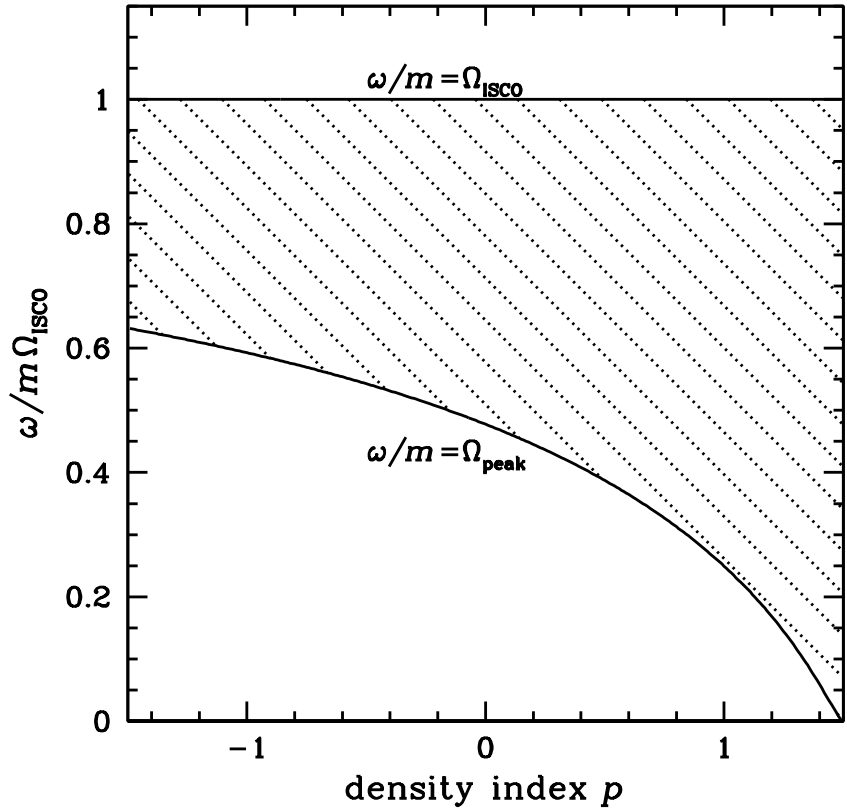}
\caption{Critical mode frequency for corotational instability as a function 
of the disc surface density index $p$ (with $\Sigma\propto r^{-p}$).
Wave absorption at the corotation resonance acts to grow the mode only if the
corotation occurs in the region of positive vortensity gradient, i.e., 
if the mode pattern frequency 
$\omega/m>\Omega_{\rm peak}$ (see Fig.~1). P-mode trapping also
requires $\omega/m<\Omega_{\rm ISCO}=\Omega(r_{\rm ISCO})$.
Note that weak mode growth can still occur when $\omega/m$ is slightly 
below $\Omega_{\rm peak}$ due to wave leakage beyond the outer Lindblad resonance
radius. See text for detail.
}
\label{fig2}
\end{figure}

\section{Calculations of Trapped, Overstable P-modes}

To determine the eigenvalues $\omega_r$ and $\omega_i$ of the trapped
modes, we solve equations (\ref{eq:sig})-(\ref{eq:uphi})
(with $u_r=0$) or equation \eqref{perturbeq1} subjected to 
appropriate boundary conditions at $r_{\rm in}$ and $r_{\rm out}$. 

\subsection{``Landau'' Integration Contour}

When solving eigenvalue problem using the standard method (e.g. the
shooting method as described in Press et al 1998), we encountered a
conundrum: For $\nu>0$, we could find both a growing mode and
a decaying mode, with almost the same $\omega_r$ but opposite
$\omega_i$.
This appears to contradict our discussion in section 3. 
This conundrum arises because our 
numerical integration is confined to the real $r$ axis. However,
analogous to Landau's analysis of wave damping in a plasma (e.g.,
Lifshitz \& Pitaevskii 1981), care must
be taken in defining appropriate contour of integration across the
corotation resonance. Indeed, at corotation, equation 
\eqref{perturbeq1} contains a singular term, proportional to 
\be
{1\over \tilde\omega} \propto {1\over r-R_c},
\ee
where $R_c \equiv r_c - ir_c\omega_i/(q\omega_r)$ is the complex
pole, $r_c$ is determined by $\omega_r=m\Omega(r_c)$ and 
$q=-(d\ln\Omega/d\ln r)_c>0$.

As discussed in Lin (1955) in the context of hydrodynamical shear flows,  
to obtain physically relevant solutions of the fluid system, it is 
necessary that the integration contour lies above the pole.
This is the Landau contour. In essence, only by adopting such
a Landau contour can one obtain the correct wave absorption (dissipation)
at the corotation. For growing modes ($\omega_i>0$), Im$(R_c)<0$,
our numerical integration along the real $r$ axis constitutes
the correct Landau contour. On the other hand, 
for decaying modes ($\omega_i<0$), the real $r$ axis is not the
correct Landau contour as Im$(R_c)>0$. Instead, to obtain physical
solutions for these decaying modes, 
the integration contour must be deformed so that $R_c$ lies
below it (see Fig.~3). As we are primarily interested in over-stable modes
in this paper, it is adequate to integrate along the real $r$ axis
in our calculation.
 

\begin{figure}
\centering
\includegraphics[width=9cm]{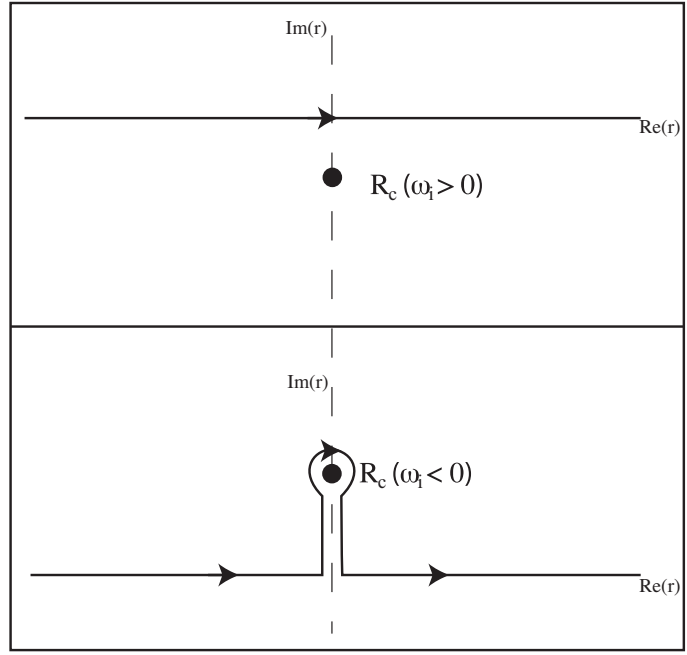}
\caption{``Landau'' contour for integration across the corotation resonance.
To calculate the growing mode ($\omega_i>0$), 
it is adequate to integrate the fluid perturbation 
equations along the real $r$ axis (upper panel). To obtain the physical solution
for the shrinking mode ($\omega_i<0$),
the integration contour must deformed so that the corotational pole 
$R_c$ lies below the contour. 
} 
\label{fig3}
\end{figure}

\subsection{Outer Boundary Condition}

As we are interested in self-excited modes in the inner region of the disc,
we adopt the radiative outer boundary condition. Specifically,
far from the outer Lindblad resonance ($r > r_{\rm OL}$) we demand that 
only an outgoing wave exists:
\be
\delta h \propto A \exp\left(i\int^r k~dr \right)~,\quad
{\rm with}~~A=\left({D\over r\Sigma k}\right)^{1/2},
\ee
where $k = \sqrt{-D/c_s^2}$ (see
Tsang \& Lai 2008a). This gives the boundary condition at some 
$r_{\rm out} > r_{\rm OL}$: 
\be
\delta h'(r_{\rm out}) 
= \delta h(r_{\rm out}) 
\left(ik + {1\over A}{dA\over dr}\right)_{r_{\rm out}}.
\label{eq:outer}\ee
In practice, we find that $r_{\rm out}\sim 2 r_{\rm OL}$ would yield sufficiently 
accurate results.

\subsection{Inner Boundary Conditions}

To obtain global trapped modes, at least partial wave reflection must occur 
at $r_{\rm in}$. To focus on the effect of corotational instability discussed
in section 3, in this section we consider two simple inner boundary 
conditions. We defer our analysis of the effect of radial inflow
on the p-modes to section 5.

(i) At the ISCO, the flow plunges into the black hole, we expect a sudden
decrease in the surface density of the disc. Thus, it is reasonable to 
consider the free surface boundary condition, i.e.,
the Lagrangian pressure perturbation $\Delta P=0$.
Using $\delta u_r=-i\tomega\xi_r$ (where $\xi_r$ is the radial Lagrangian
displacement), and $\Delta P=\delta P+\xi_r dP/dr$, we have
\be
\left({\Delta P\over\Sigma}\right)_{\rm r_{\rm in}}=
\left(\delta h - pc_s^2\frac{i\delta u_r}{r \tomega}\right)_{r_{\rm 
in}}=0,
\label{eq:DeltaP}\ee 
where we have used
$dP/dr = (dP/d\Sigma)(d\Sigma/dr) = -pc_s^2\Sigma/r$
for barotropic, power-law discs ($\Sigma\propto r^{-p}$).

(ii) We assume that 
the radial velocity perturbation vanishes at the inner boundary, i.e., 
$\delta u_r = 0$. This was adopted by Tagger \& Varniere (2006)
in their calculations of overstable global modes due to accretion-ejection 
instability.

Both of these boundary conditions correspond to zero loss of wave energy
at the inner boundary: If a wave from large radii impinges toward 
$r_{\rm in}$, the reflected wave will have the same amplitude.
However, the phase shifts due to reflection
differ in the two cases, and the resulting mode frequencies $\omega_r$ 
are different. Since the corotational wave amplification depends on
$\omega_r$, the mode growth rate $\omega_i$ will also be different.

\subsection{Numerical Results}

\begin{figure}
\centering
\begin{tabular}{ccc}
\epsfig{file=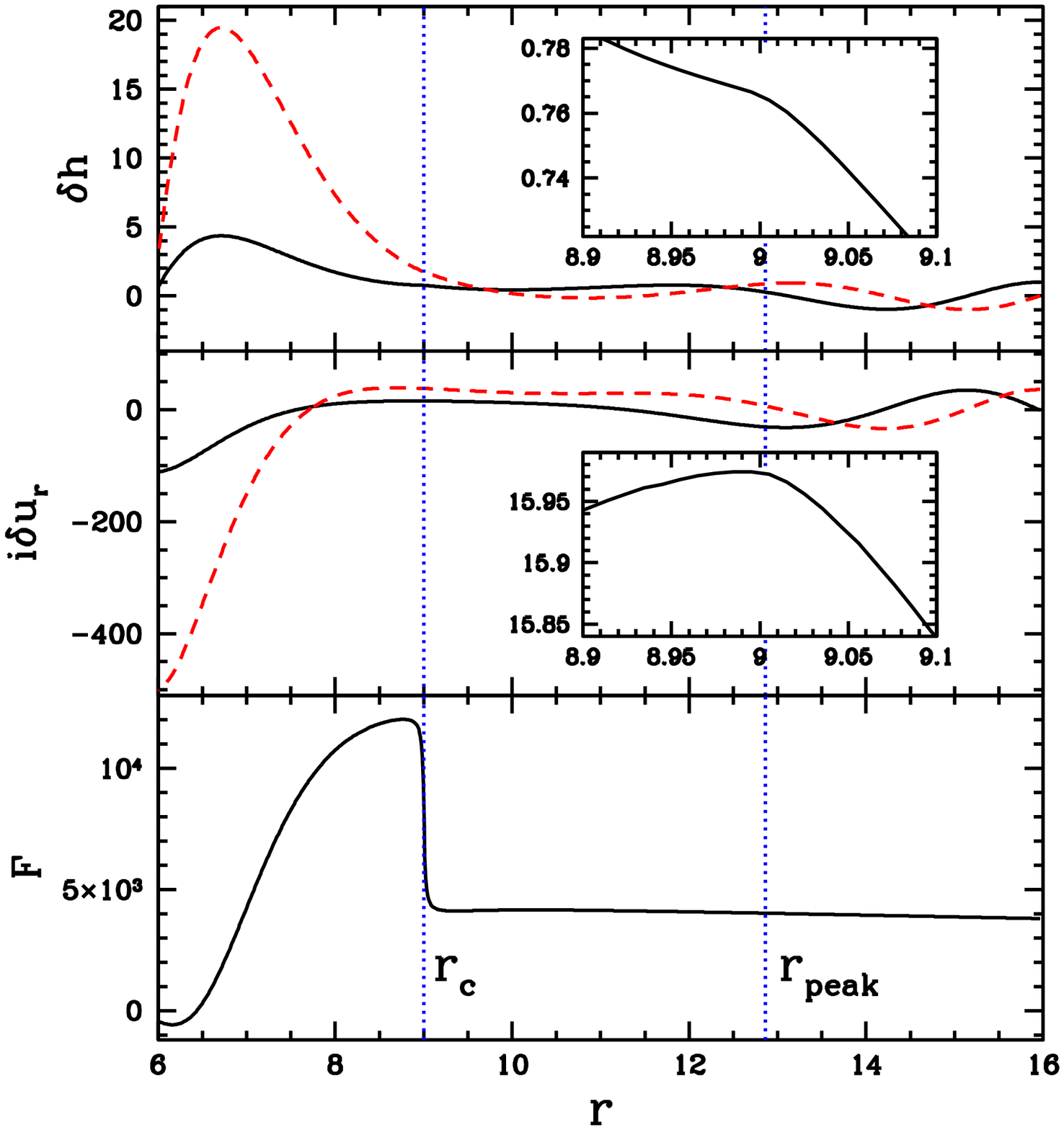,width=0.5\linewidth,clip=} &
\epsfig{file=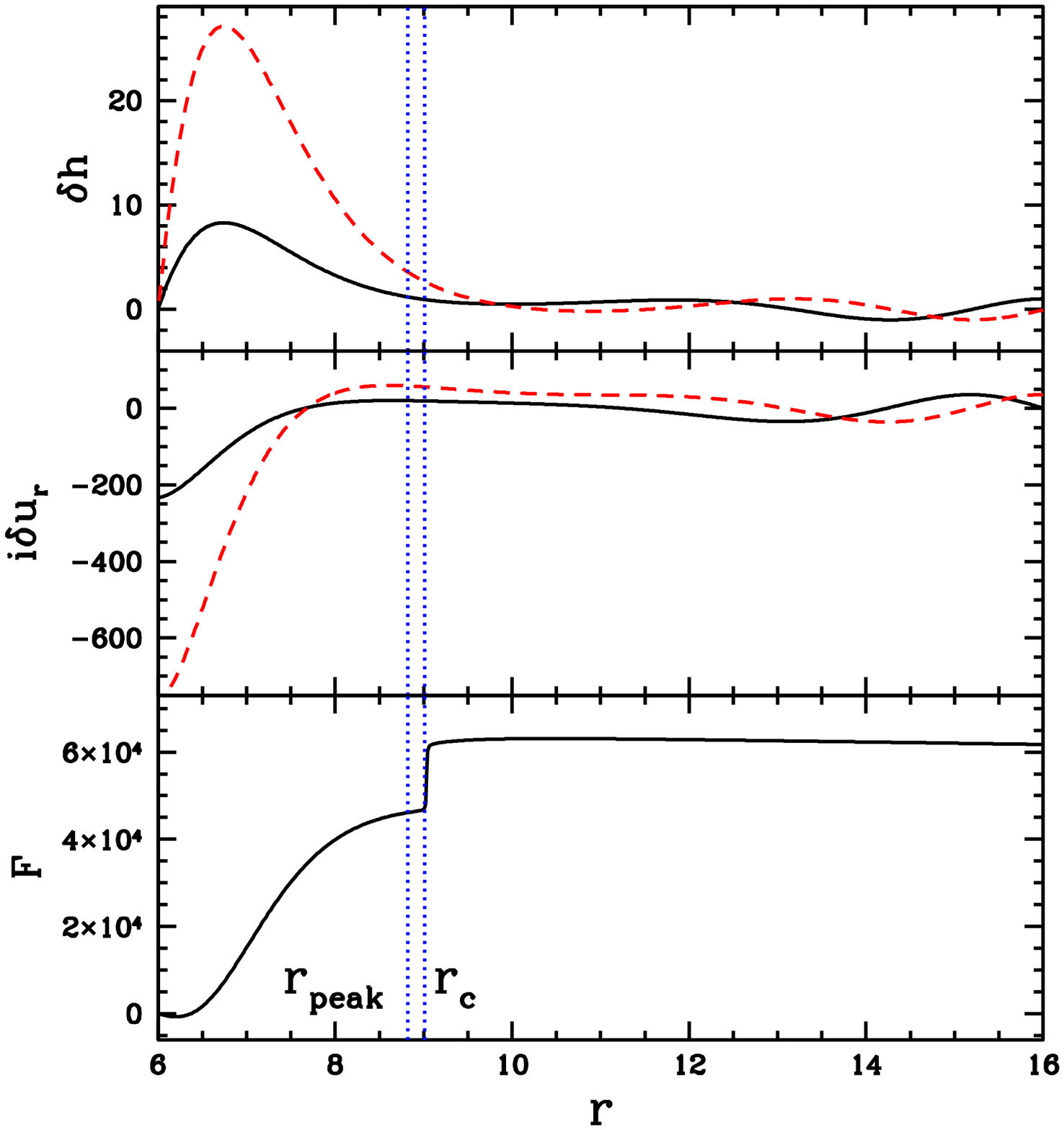,width=0.5\linewidth,clip=}
\end{tabular}
\caption{Example wavefunctions for disc p-modes. The upper and 
middle panels show $\delta h$ and $i\delta u_r$ (the solid lines for
the real part and dashed lines for the imaginary part),
the lower panels show the angular momentum flux, all in arbitrary 
units [with $\delta h(r_{\rm out})=1$]. The radius $r$ is in units of
$GM/c^2$. The disc sound speed is $c_s =0.1r\Omega$, and the $m=2$ 
modes are obtained using the inner boundary condition $\Delta 
P(r_{\rm ISCO}) = 0$. 
The left panels show the p-mode for the disc with 
a density profile $\Sigma \propto r^{-1}$, with 
the eigenvalues $\omega_r = 0.467 m\Omega_{\rm ISCO},~ \omega_i/\omega_r
= 0.0029$ [where $\Omega_{\rm ISCO}
=\Omega(r_{\rm ISCO})$]; the right panels show the mode for 
the disc with constant surface density profile ($p=0$), with
eigenvalues $\omega_r =0.464 m\Omega_{\rm ISCO}, ~
\omega_i/\omega_r = 0.00073$. 
Note that the model shown on the left panels has $r_c<r_{\rm peak}$
(the radius of peak vortensity) and thus $F(r_c+)<F(r_c-)$, 
while the model shown on the right panels has $r_c>r_{\rm peak}$
and $F(r_c+)>F(r_c-)$.
In both models there is a positive flux for $r > r_c$ due to the outward
propagating wave. The inserts on the left panels show the blowups
of the real wavefunctions near the corotation radius.
}
\label{fig4}
\end{figure}

\begin{figure}
\centering
\begin{tabular}{ccc}
\epsfig{file=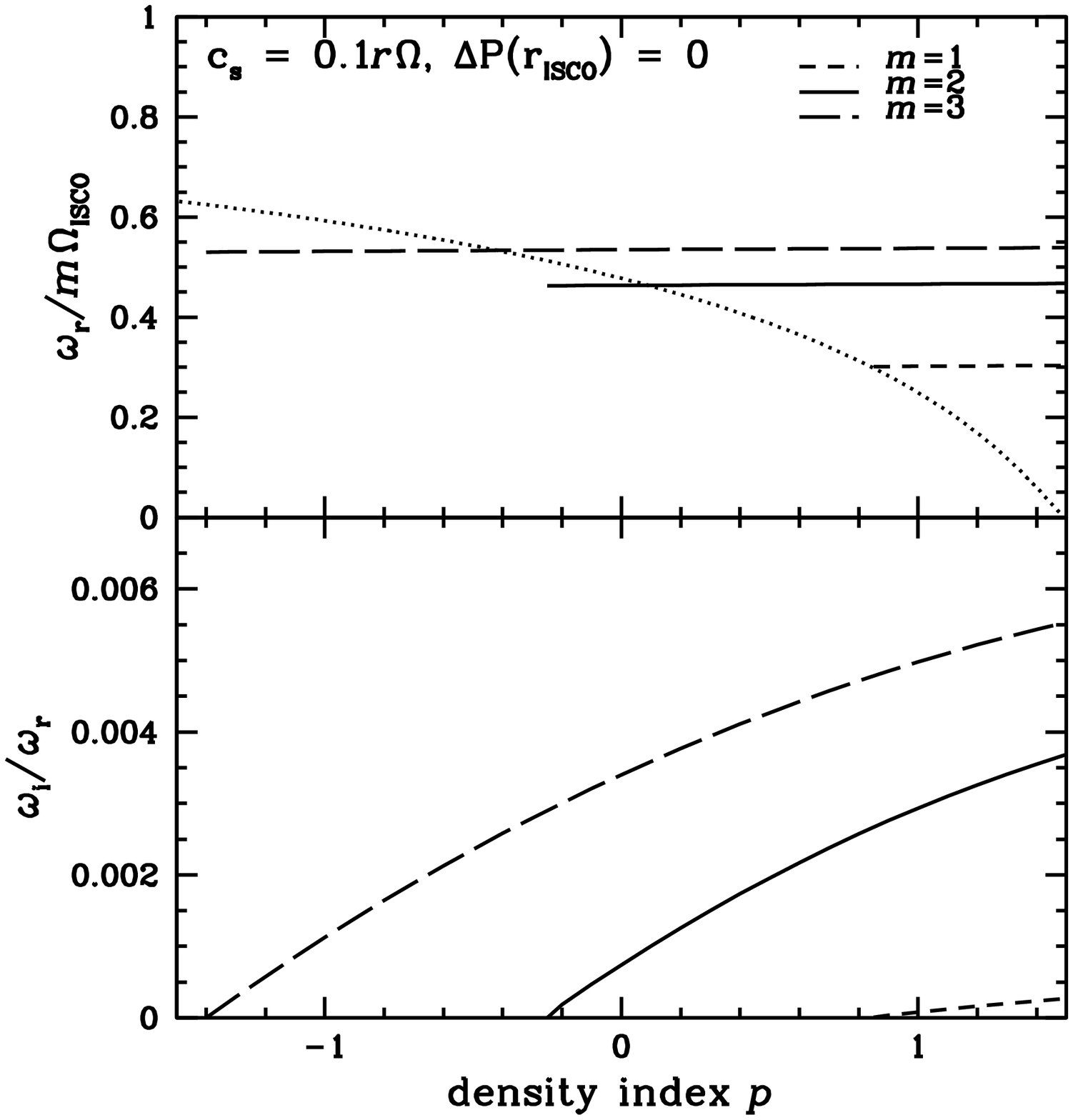,width=0.5\linewidth,clip=} &
\epsfig{file=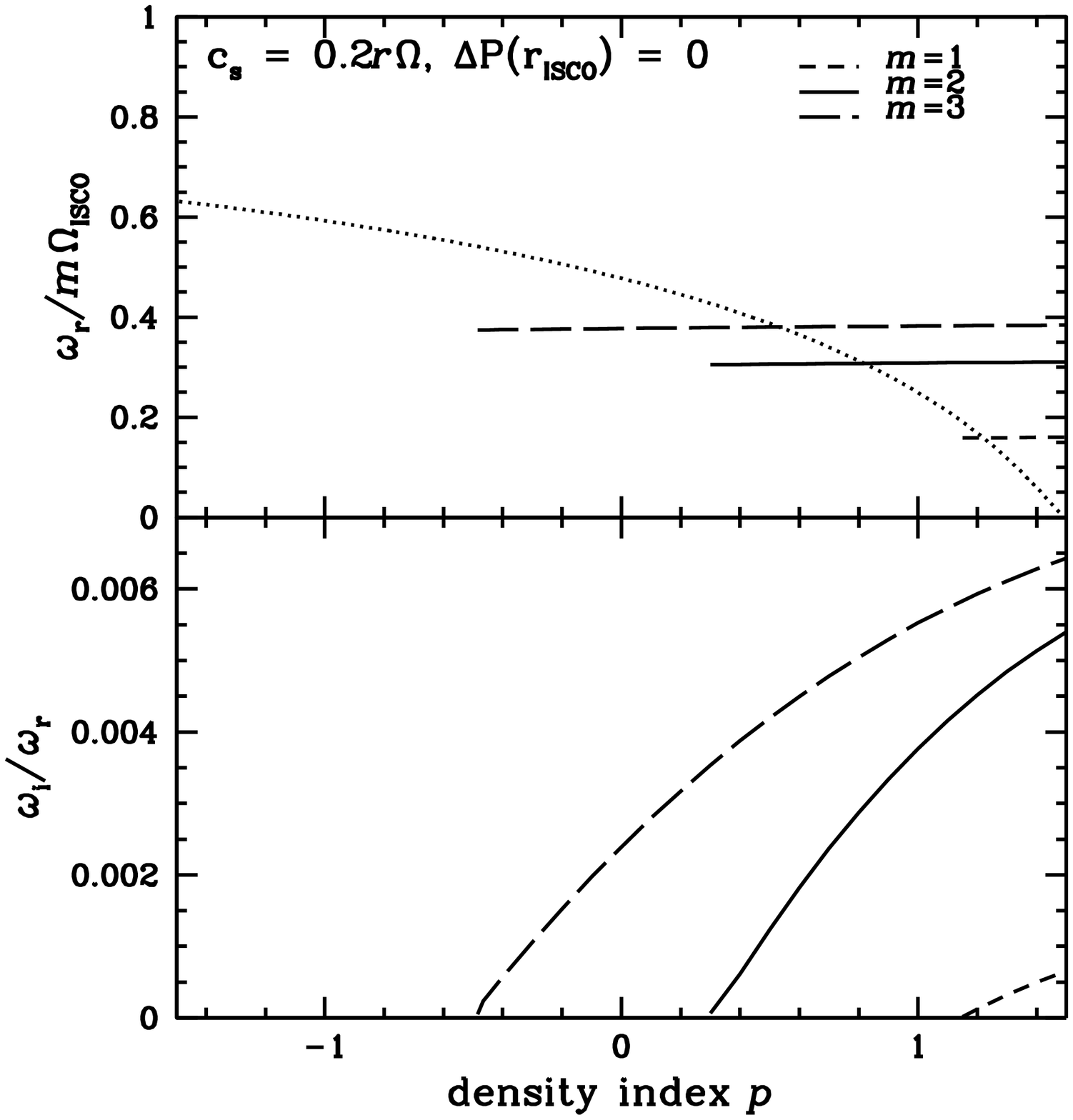,width=0.5\linewidth,clip=}
\end{tabular}
\caption{The real and imaginary frequencies of disc p-modes (with 
azimuthal wave numbers $m=1,2,3$)
as a function of the surface density index $p$ (where $\Sigma
\propto r^{-p}$). The modes are calculated assuming the inner 
boundary condition $\Delta P(r_{\rm ISCO}) = 0$.
The left panels are for discs with $c_s=0.1r\Omega$ and the right
panels for $c_s=0.2r\Omega$. 
The dotted lines denote the lower bound $\omega_r/m = \Omega_{\rm
peak}$ for which the corotational wave absorption 
acts to enhance mode growth.}
\label{fig5}
\end{figure}

\begin{figure}
\centering
\epsfig{file=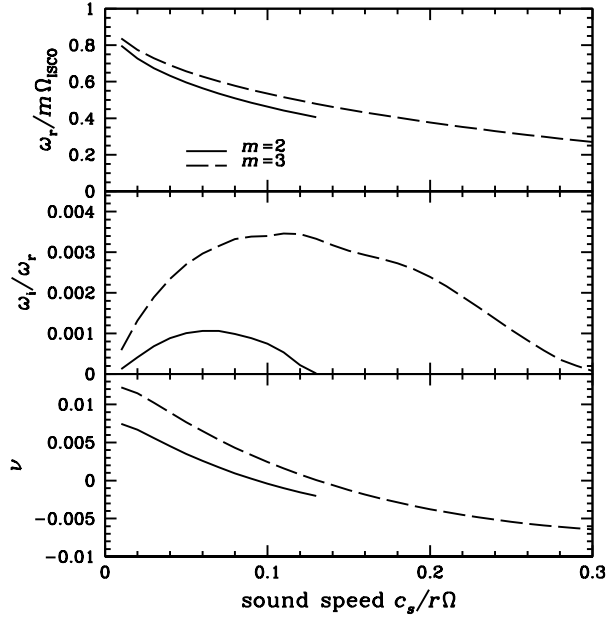,width=0.5\linewidth,clip=}
\caption{The real and imaginary frequencies of disc p-modes
(with $m=2,3$) as a function of the normalized sound speed
$c_s/(r\Omega)$. The disc is assumed to have a constant
density profile ($p = 0$), and the inner boundary condition
is $\Delta P(r_{\rm ISCO}) = 0$. The bottom panel shows
the $\nu$ parameters for the modes.}
\label{fig6}
\end{figure}

\begin{figure}
\centering
\begin{tabular}{ccc}
\epsfig{file=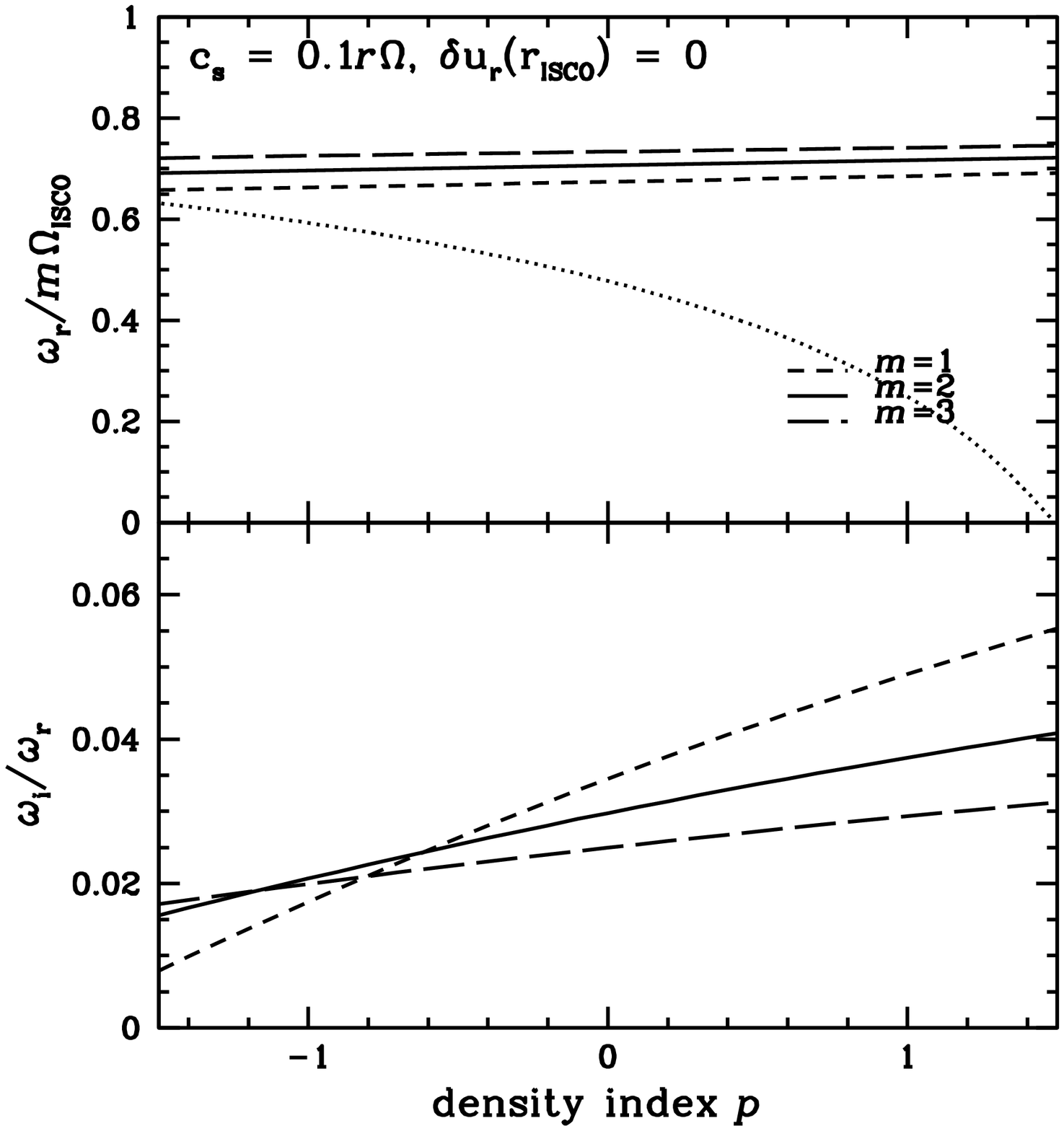,width=0.5\linewidth,clip=} &
\epsfig{file=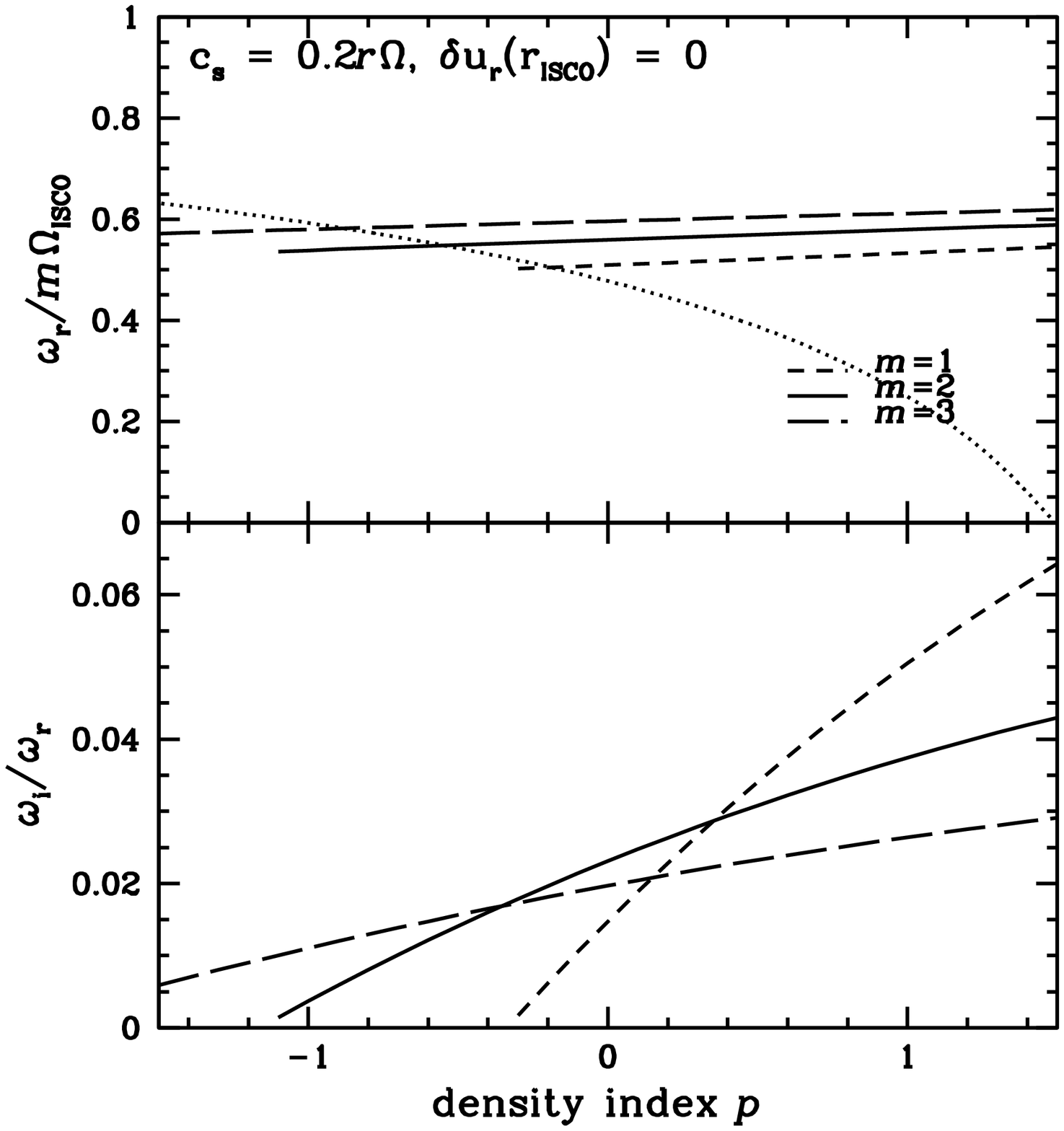,width=0.5\linewidth,clip=}
\end{tabular}
\caption{The real and imaginary frequencies of disc p-modes (with 
azimuthal wave numbers $m=1,2,3$)
as a function of the surface density index $p$ (where $\Sigma
\propto r^{-p}$). The modes are calculated assuming the inner 
boundary condition $\delta u_r(r_{\rm ISCO}) = 0$.
The left panels are for discs with $c_s=0.1r\Omega$ and the right
panels for $c_s=0.2r\Omega$. 
The dotted lines denote the lower bound $\omega/m = \Omega_{\rm
peak}$ for which the corotational wave absorption 
acts to enhance mode growth.}
\label{fig7}
\end{figure}

We solve for the complex eigen-frequency $\omega = \omega_r + i
\omega_i$ using the shooting method, with a 
fifth-order Runge-Kutta integrator (Press et al.~1992). As
discussed in section 4.1 we only calculate the growing modes ($\omega_i>0$). We
consider disc models with different surface density profile (characterized
by the index $p$), sound speed $c_s$, and inner boundary
conditions. For a given set of disc parameters 
and the azimuthal mode wavenumber $m$, the 
lowest order (highest frequency) mode has the best chance
of being overstable. This is easily understood from our discussion in
section 3 (see Fig.~1): a low-frequency wave has to penetrate a wider 
evanescent barrier for the corotational amplifier to be effective,
and when $\omega_r<m\Omega_{\rm peak}$ the corotation resonance acts to
damp the mode. For most disc models we have considered, 
the lowest order mode (of a given $m$)
is the only mode that has $\omega_i>0$.

Our numerical results are presented in Figures \ref{fig4}--\ref{fig7}.
Figure 4 gives two examples of the eigenfunctions of overstable trapped 
p-modes, obtained with the inner boundary condition $\Delta P=0$.
In addition to $\delta h$ and $\delta u_r$, we also plot the
angular momentum flux carried by the wave across the disc
(e.g., Goldreich \& Tremaine 1979; Zhang \& Lai 2006)
\be
F(r) = \pi r^2\Sigma\, {\rm Re} \left(\delta u_r \delta u_\phi^\ast\right)=
\frac{\pi m r \Sigma}{D}\,{\rm Im}\left(\delta h\frac{d\delta
h^*}{dr} \right),
\label{eq:flux}\ee
where the second equality follows from equations 
(\ref{eq:sig})-(\ref{eq:uphi}) (with $u_r=0$).
We see from Fig.~4 that outside the corotation radius ($r_c$), 
$F$ is nearly constant since only
the outgoing wave exists in this region and the wave action is conserved
in the limit of $\omega_i\ll \omega_r$. Inside $r_c$, the interference between
the ingoing and outgoing waves gives rise to the variation of $F$. At 
$r_{\rm ISCO}$, $F$ approaches zero since no wave action is lost through 
the disc inner boundary when $\Delta P=0$.\footnote{
Note that equation (\ref{eq:flux}) is the time averaged flux and is defined
for waves with real $\omega$.
Using equation (\ref{eq:DeltaP}) and 
$D\delta u_r=i\tomega d\delta h/dr-(2im\Omega/r)\delta h$ (obtained from 
eqs.~[\ref{eq:ur}]-[\ref{eq:uphi}]) it is easy to
show that $F(r_{\rm in})=0$ exactly for real $\omega$.}
Figure 4 also shows a flux jump across the corotation.
In the limit of $\omega_i\ll \omega_r$, $d\delta h/dr$,
$\delta u_r$ and $\delta u_\phi$
are discontinuous across $r_c$
(although $\delta h$ is continuous), giving rise to the flux discontinuity
(see Tsang \& Lai 2008a):
\be
F(r_c+)-F(r_c-)=-{2\pi^2 m r\Sigma \nu \kappa\over c_s D}|\delta h|^2
\Bigl|_{r_c}.
\ee
This discontinuity signifies the corotational wave absorption, the
sign of which depends on $\nu\propto (d\zeta/dr)_{r_c}$, as discussed in
section 3. Thus, the model
shown on the left panels of Fig.~4 has $r_c<r_{\rm peak}$ and
$\nu>0$, and the mode growth is primarily driven by wave absorption 
at the corotation. The model shown on the right panels of Fig.~4, 
on the other hand, has $r_c>r_{\rm peak}$ and $\nu<0$, thus the corotational
wave absorption acts to damp the mode, and the overall growth of the mode
is due to the outgoing wave beyond $r_c$, and the growth rate 
is much smaller than the model shown on the left panels.

Figures 5--6 show the frequencies of the
fundamental (no node/highest frequency) growing 
p-modes (with $m=1,2,3$) for various disc parameters, again obtained with 
the inner boundary condition $\Delta P=0$. We consider 
$p$ in the range between $-1.5$ and 1.5, and $c_s$ 
up to $0.3r\Omega$. 
For a given sound speed,
the real mode frequency $\omega_r$ depends very weakly on $p$, but
the growth rate $\omega_i$ increases with $p$ (see Fig.~5) since
a larger value of $p$ leads to a larger $\nu$ and enhanced wave
absorption at the corotation [see equation (\ref{eq:nu})].
In general, as the sound speed 
increases, the effective wavelength of the mode increases, and 
$\omega_r$ decreases in order ``fit in'' the trapping zone between
$r_{\rm in}$ and $r_{\rm IL}$ (see Fig.~6). 
The mode growth rates $\omega_i$ depends 
on $c_s$ in a non-monotonic way because of two competing effects:
As $c_s$ increases, less attenuation occurs in the evanescent zone, 
and more wave energy can be absorbed at the corotation and
propagate to the outer edge of the disc; these tend to increase
$\omega_i$. On the other hand, increasing $c_s$ also leads to
smaller $\omega_r$, which shifts the corotation resonance to 
a larger radius and leads to decreasing $\nu$ and $\omega_i$.

Note that the growing modes shown in Fig.~5 extend below the 
$\omega_r/m=\Omega_{\rm peak}$ boundary due to the propagation of 
waves beyond the corotation radius, as discussed in section 3
[see eqs.~(\ref{eq:nucrit})].
Such modes (with $\nu_{\rm crit}<\nu<0$) grow
significantly slower than the modes with $\nu>0$ 
as the flux is attenuated by the entire barrier between $r_{\rm IL}$ and $r_{\rm OL}$.

For comparison, Figure 7 shows the disc mode
frequencies and growth rates
when the inner boundary condition $\delta u_r=0$ is
adopted. The different boundary condition leads to a different phase shift
$\varphi$ and higher mode frequency, but 
the results are similar to those illustrated in Fig.~5.
In particular, as $p$ increases, $\omega_r$ remains approximately
constant while $\omega_i$ increases.

\section{Effect of Radial Inflow on the P-Mode Growth Rate}

Our mode calculations presented in section 4 neglect 
the radial velocity of the accretion flow and assume
a loss-less inner disc boundary condition (either $\Delta P=0$ or 
$\delta u_r=0$ at $r_{\rm ISCO}$). In real discs, the radial 
inflow velocity $u_r$ is not negligible as $r$ approaches 
$r_{\rm ISCO}$, and the flow goes through a transonic point (where
$u_r=-c_s$) at a radius very close to $r_{\rm ISCO}$. We expect 
that part of the fluid perturbations may be advected into the
black hole and the inner disc boundary will not be completely
loss-less. Here we study the effect of the transonic flow
on the p-mode growth rate.

We note that just as the accretion disc is not laminar but 
turbulent, the accretion flow around $r_{\rm ISCO}$
is complicated. General relativistic MHD simulations in 3D
are only beginning to shed light on the property of the black hole
accretion flow (e.g., Beckwith, Hawley \& Krolik 2008; Shafee et al.~2008;
Noble, Krolik \& Hawley 2008), and many uncertainties remain 
unresolved. Here, to make analytic progress, we adopt a simple viscous
transonic flow model, which qualitatively describes the inner accretion flow
of the black hole as long as the flow remains 
geometrically thin (see Afshordi \& Paczynski 2003).

\subsection{Boundary Condition at the Sonic Point}

We rewrite equations (\ref{eq:sig})-(\ref{eq:ur}) as
\ba
&&{u_r\over c_s^2}\delta h'+\delta u_r'=\left[{i\tomega\over c_s^2}
-u_r\left(c_s^{-1}\right)'\right]\delta h+{u_r'\over u_r}\,\delta u_r-{im\over r}
\,\delta u_\phi\equiv {\cal A}_1,\\
&&\delta h'+u_r\,\delta u_r'=(i\tomega-u_r')\,\delta u_r+2\Omega\,\delta u_\phi
\equiv {\cal A}_2,
\ea
where $'$ stands for $d/dr$ and we have used $r\Sigma u_r=$ constant for 
the background flow. Solving for $\delta h'$ and $\delta u_r'$ we have
\ba
&&\delta h'={{\cal A}_2-{\cal A}_1u_r\over 1-u_r^2/c_s^2},\\
&&\delta u_r'={{\cal A}_1-(u_r/c_s^2){\cal A}_2 \over 1-u_r^2/c_s^2}.
\ea
Clearly, in order for the perturbation to be regular at the sonic
point $r_s$, where $u_r=-c_s$, we require
\be
{\cal A}_2+c_s{\cal A}_1=0\quad {\rm at}~~r=r_s.
\label{eq:calA}\ee
For definiteness, we characterize the variations of $\Sigma$ and $c_s$ at the
sonic point $r_s\simeq r_{\rm ISCO}$ by the two length scales: 
\be
\left({\Sigma'\over \Sigma}\right)_{r_s}={1\over L_\Sigma},\qquad
\left({c_s'\over c_s}\right)_{r_s}={1\over L_c}.
\label{eq:Lc}\ee
From $r\Sigma u_r=$ constant, we also have $u_r'=c_s (r_s^{-1}+L_\Sigma^{-1})$ at $r=r_s$.
Then equation (\ref{eq:calA}) becomes
\be
\left({i\tomega\over c_s}-{2\over L_c}\right)\delta h
+\left[i\tomega-2c_s\left({1\over r}+{1\over L_\Sigma}\right)\right]\delta u_r
+\left(2\Omega-{imc_s\over r}\right)\delta u_\phi=0\quad {\rm at}~~r=r_s.
\label{eq:bc}\ee
This is the boundary condition for the fluid perturbations at the sonic point.

\subsection{Properties of the Transonic Flow}

Before exploring the effect the radial inflow on the disc modes, we first 
estimate the length scale for the surface density variation, 
$L_\Sigma$, using the viscous slim disc model (e.g.,
Muchotrzeb \& Paczynski 1982; Matsumoto et al.~1984;
Abramowicz et al.~1988)

The basic steady-state slim disc equations are
\ba
&&\dot M=-2\pi r\Sigma u_r,\\
&&u_ru_r'=-c_s^2{\Sigma'\over\Sigma}+(\Omega^2-\Omega_K^2)r,\\
&&{\dot M} l_0={\dot M} l+2\pi\nu_{\rm vis} r^3\Sigma\Omega',\label{eq:l0}
\ea
where $l=r^2\Omega$ is the specific angular momentum of the flow, 
$\Omega$ is the actual rotation rate,
$\Omega_K$ is given by equation (\ref{eq:OmegaK}), $\nu_{\rm vis}$ 
is the kinetic viscosity,
and $l_0$ is the eigenvalue that must be solved so that flow pass through the 
sonic point smoothly. We shall use the $\alpha$-disc model,
so that $\nu_{\rm vis}=\alpha Hc_s$, with $H\simeq c_s/\Omega_K$.

To estimate $L_\Sigma$, we assume $l(r)\simeq r^2\Omega_K(r)$ for $r\go r_{\rm ISCO}$
and $l_0\simeq l_K(r_{\rm ISCO})$. Equation (\ref{eq:l0})
gives 
\be
u_r(1-l_0/l)\simeq -3\nu_{\rm vis}/(2r)=-3\alpha Hc_s/(2r),
\ee
valid for $r\go r_{\rm ISCO}$. At the radius $r=r_{\rm ISCO}+\Delta r$, we have
$u_r(\Delta r/r_{\rm ISCO})^2\simeq -4\alpha Hc_s/r_{\rm ISCO}$.
The sonic point ($u_r=-c_s$) is at $\Delta r\simeq 2\sqrt{\alpha Hr_{\rm ISCO}}$,
and $u_r=-c_s/2$ at $\Delta r\simeq 2\sqrt{2\alpha Hr_{\rm ISCO}}$. Thus
$u_r'(r_s)\sim c_s/L_\Sigma$, with 
\be
L_\Sigma=\left({\Sigma\over\Sigma'}\right)_{r_s}\sim
2\sqrt{\alpha Hr_s}=2\sqrt{\alpha\beta}\,r_s,
\ee
where $\beta=c_s/(r\Omega)$. This should be compared to the disc thickness
$H=\beta r$: depending on the value of $\alpha$, both $L_\Sigma<H$ and $L_\Sigma>H$ are
possible.

The value and sign of $L_c$ depend on the thermodynamical and radiative
properties of the flow, and cannot be estimated in a simple way.
It is reasonable to expect $|L_c|\sim r_s$.

\subsection{Reflectivity at the Sonic Point}

\begin{figure}
\centering
\begin{tabular}{ccc}
\epsfig{file=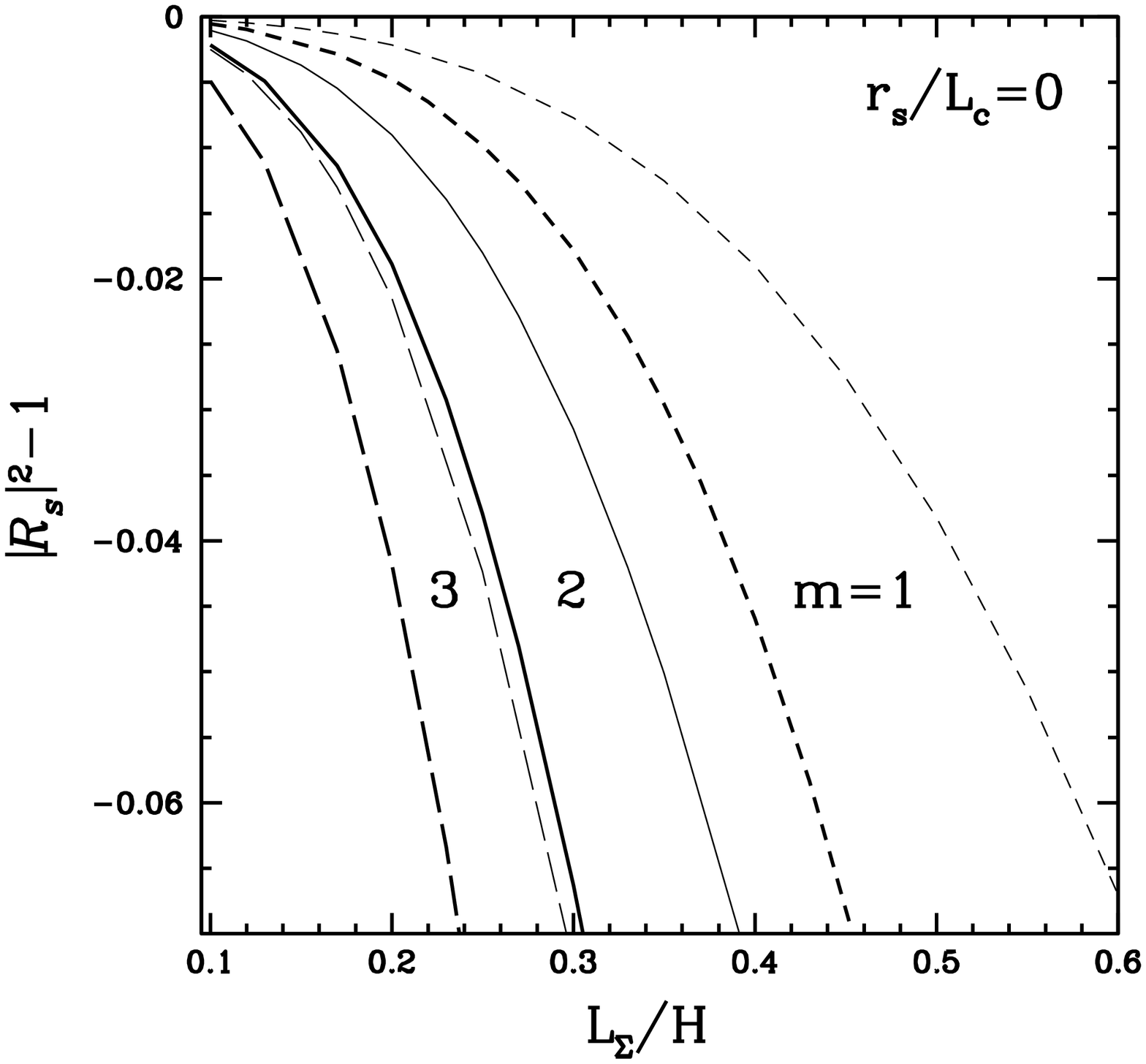,width=0.5\linewidth,clip=} &
\epsfig{file=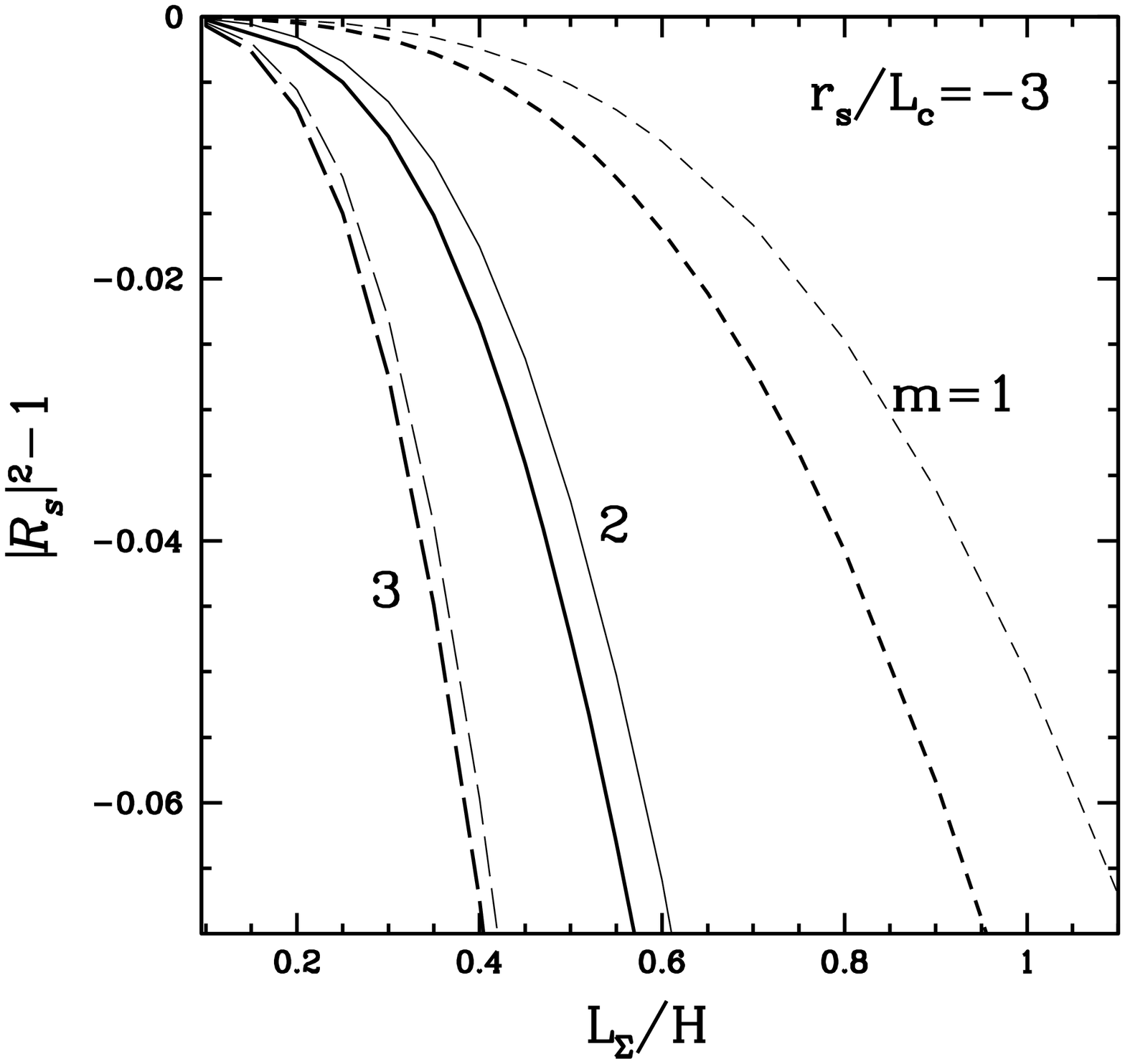,width=0.5\linewidth,clip=}
\end{tabular}
\caption{The wave reflectivity at the transonic point of the inner disc
as a function of the parameter $L_\Sigma/H$, for $r_s/L_c=0$ (the left panel)
or $r_s/L_c=-3$ (the right panel). In both panels, the heavier lines are for
$c_s=0.1r\Omega$ and the lighter lines for $c_s=0.05r\Omega$. The 
short-dashed, solid and long-dashed lines are for $m=1,2,3$, respectively.
The wave frequency is set to be $\omega=0.7m\Omega(r_s)$.
}
\label{fig8}
\end{figure}

We can understand qualitatively the effect of the transonic boundary
condition on the p-mode by calculating the reflectivity $\cR_s$
of the inner boundary.

Consider a density wave $\delta h\propto \exp(i\int^r k\,dr)$
in the wave zone $r_{\rm in}=r_s<r<r_{\rm IL}$, traveling toward
the inner disc boundary.
\footnote{Note that since the group velocity 
of the wave has opposite sign as the phase velocity for $r<r_{\rm IL}$,
the wave of the form $\exp(i\int^r k\,dr)$ (with $k>0$) is inward propagating.}. 
Upon reflection,
the wave becomes $\delta h\propto \cR_s\exp(-i\int^r k\,dr)$. 
Including the correct WKB amplitude (see Tsang \& Lai 2008a),
the wave outside the sonic point can be written as (up to a constant 
prefactor)
\be
\delta h=A\left[\exp\left(i\int^r_{r_s}k\,dr\right)
+\cR_s \exp\left(-i\int^r_{r_s} k\,dr\right)\right],
\qquad (r_s<r<r_{\rm IL})
\label{eq:deltah}\ee
where 
\be
k={(-D)^{1/2}\over c_s},\quad
A=\left({D\over r\Sigma k}\right)^{1/2}.
\ee
To apply the boundary condition (\ref{eq:bc})
to equation (\ref{eq:deltah}), we neglect
$u_r$ in equations (\ref{eq:sig})--(\ref{eq:uphi}) at $r=r_s+\varepsilon$,
with $\varepsilon\ll r_s$ and $r_s\simeq r_{\rm ISCO}$. Implicit in
this procedure is the assumption that the fluid perturbations
do not vary significantly between $r_s$ and $r_s+\varepsilon$. 
We then obtain
\be
\cR_s={ik+L_A^{-1}+K-2m\Omega/(r\tomega)\over
ik-L_A^{-1}-K+2m\Omega/(r\tomega)}\Biggl|_{r_s},
\ee
where
\be 
L_A^{-1}=\left({A'\over A}\right)_{r_s},
\ee
and
\be
K=\left({\tomega^2L_\Sigma\over 2c_s^2}\right)
{1-\left({mc_s/r\tomega}\right)^2-i\left[{2mc_s\Omega/ 
(r\tomega^2)}-{2c_s/(\tomega L_c)}\right]\over
1+({L_\Sigma/r})-i({\tomega L_\Sigma/2c_s})}\Biggl|_{r_s}.
\label{eq:K}\ee

When considering the damping of the p-mode 
due to the transonic flow, the quantity
$|\cR_s|^2-1$ is the most relevant (see section 5.4 below). Let 
\be
K=|K|\exp(-i\psi),
\ee
we have
\be
|\cR_s|^2-1=-{4\,k\,|K|\sin\psi\over (L_A^{-1}+|K|\cos\psi)^2
+(k+|K|\sin\psi)^2}\Biggl|_{r_s}.
\ee
Using $\beta =c_s/(r\Omega)$, $\tomega=-\hat\omega m\Omega_{\rm ISCO}$ (where
$0<{\hat\omega}<1$), we find from equation (\ref{eq:K}) 
that 
\be
\psi=\tan^{-1}\left({2\beta\over m{\hat\omega}^2}{1+r_s\hat\omega/L_c
\over 1-\beta^2/\hat\omega^2}\right)
+\tan^{-1}\left({m\hat\omega L_\Sigma\over 2\beta r_s}{1\over 1+L_\Sigma/r_s}
\right).
\ee

Figure 8 shows how the reflectivity depends on various parameters 
of the disc inner edge. In particular, for small $L_\Sigma/H=L_\Sigma/(\beta r_s)$,
i.e., when the surface density of the disc decreases rapidly at the sonic point, 
$|\cR_s|^2$ is only slightly smaller than unity and the wave loss
at the inner edge of the disc is small.

\subsection{Mode Growth Rate in the WKB Approximation}

Consider the p-mode trapped between $r_{\rm in}=r_s\simeq r_{\rm ISCO}$ and
$r_{\rm IL}$. With the reflectivity
at $r_{\rm IL}$ given by $\cR$ (see section 3), we can write the wave amplitude 
for $r<r_{\rm IL}$ as\footnote{Note that this definition of $\cR$
differs from that in Tsang \& Lai (2008a) by a phase factor of
$\exp(i\pi/4)$.}
\be
\delta h \propto \left({D\over r\Sigma k}\right)^{1/2}\left[
\exp\left(-i\int^r_{r_{\rm IL}}k\,dr\right)
+\cR \exp\left(i\int^r_{r_{\rm IL}} k\,dr\right)\right],
\qquad (r_s<r<r_{\rm IL})
\label{eq:deltah2}\ee
On the other hand, with the reflectivity at 
$r_{\rm in}=r_s$ given by $\cR_s$, the wave can also be expressed as
(\ref{eq:deltah}). For stationary waves we therefore require
\be
\exp(2i\Theta)=\cR\cR_s,\qquad {\rm with}~~\Theta=\int_{r_{\rm in}}^{r_{\rm IL}}k\,dr
=\Theta_r+i\Theta_i,
\ee
where $\Theta_r$ and $\Theta_i$ are real. The real eigen-frequency 
$\omega_r$ is given by
\be
\Theta_r=\int_{r_{\rm in}}^{r_{\rm IL}}k_r\,dr=
\int_{r_{\rm in}}^{r_{\rm IL}}{\sqrt{\tomega_r^2-\kappa^2}\over c_s}\,dr=
n\pi+{\varphi\over 2},
\ee
where $\cR\cR_s=|\cR\cR_s|\exp(i\varphi)$, and $n$ is an integer. The mode growth
rate $\omega_i$ is determined by
$|\cR\cR_s|=\exp(-2\Theta_i)$, or
\be
\tanh\Theta_i=-\left({|\cR\cR_s|-1\over |\cR\cR_s|+1}\right).
\ee
For $\Theta_i=\int_{r_{\rm in}}^{r_{\rm IL}}k_i\,dr\ll 1$ and
$k_i\simeq \omega_i\tomega_r/(c_s\sqrt{\tomega_r^2-\kappa^2})$, we obtain
\be
\omega_i = \left(\frac{|\cR\cR_s| -1}{|\cR\cR_s|+1}\right)
\left[\int_{r_{\rm in\ }}^{r_{\rm IL}}
{|\tomega_r|\over c_s\sqrt{\tomega_r^2 - \kappa^2}} dr\right]^{-1},  
\label{eq:omegai2}\ee
where we have assumed $\tomega_r<0$.
Equation (\ref{eq:omegai2}) is to be compared with (\ref{eq:omegai}), 
where perfect reflection at $r_{\rm in}$ is assumed.
Clearly, to obtain growing modes we require $|\cR\cR_s|>1$. For a given
$|\cR|>1$, growing modes are possible only when the loss
at the sonic point is sufficiently small (i.e., $|\cR_s|$ is sufficiently
close to unity).

\subsection{Numerical Results}

\begin{figure}
\centering
\epsfig{file=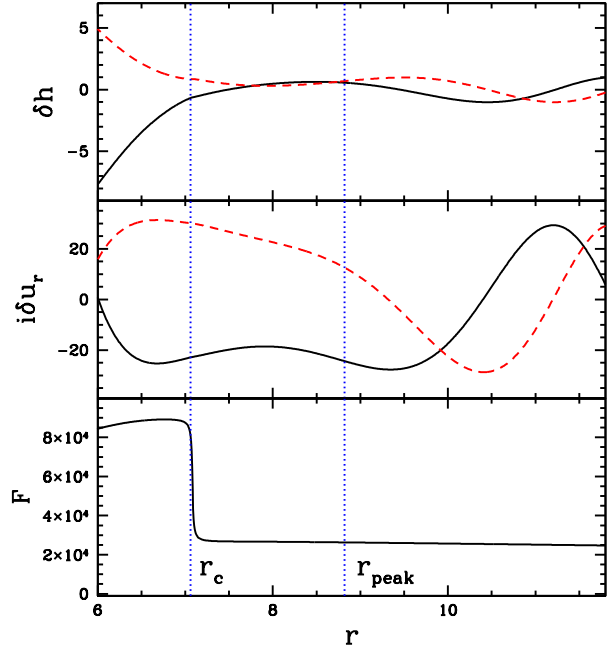,width=0.5\linewidth,clip=}
\caption{Wavefunctions for a disc p-mode. The 
notations are the same as in Fig.~4.
The disc has sound speed $c_s =0.1r\Omega$ and constant density profile
($p=0$), and the $m=2$ mode is calculated using 
the transonic inner boundary condition (\ref{eq:bc}) with 
$L_\Sigma/H=0.25$ and $L_c=\infty$.
The eigenvalues are $\omega_r = 0.725m\Omega_{\rm ISCO}$ and
$\omega_i/\omega_r=0.00267$.
}
\label{fig9}
\end{figure}

\begin{figure}
\centering
\begin{tabular}{ccc}
\epsfig{file=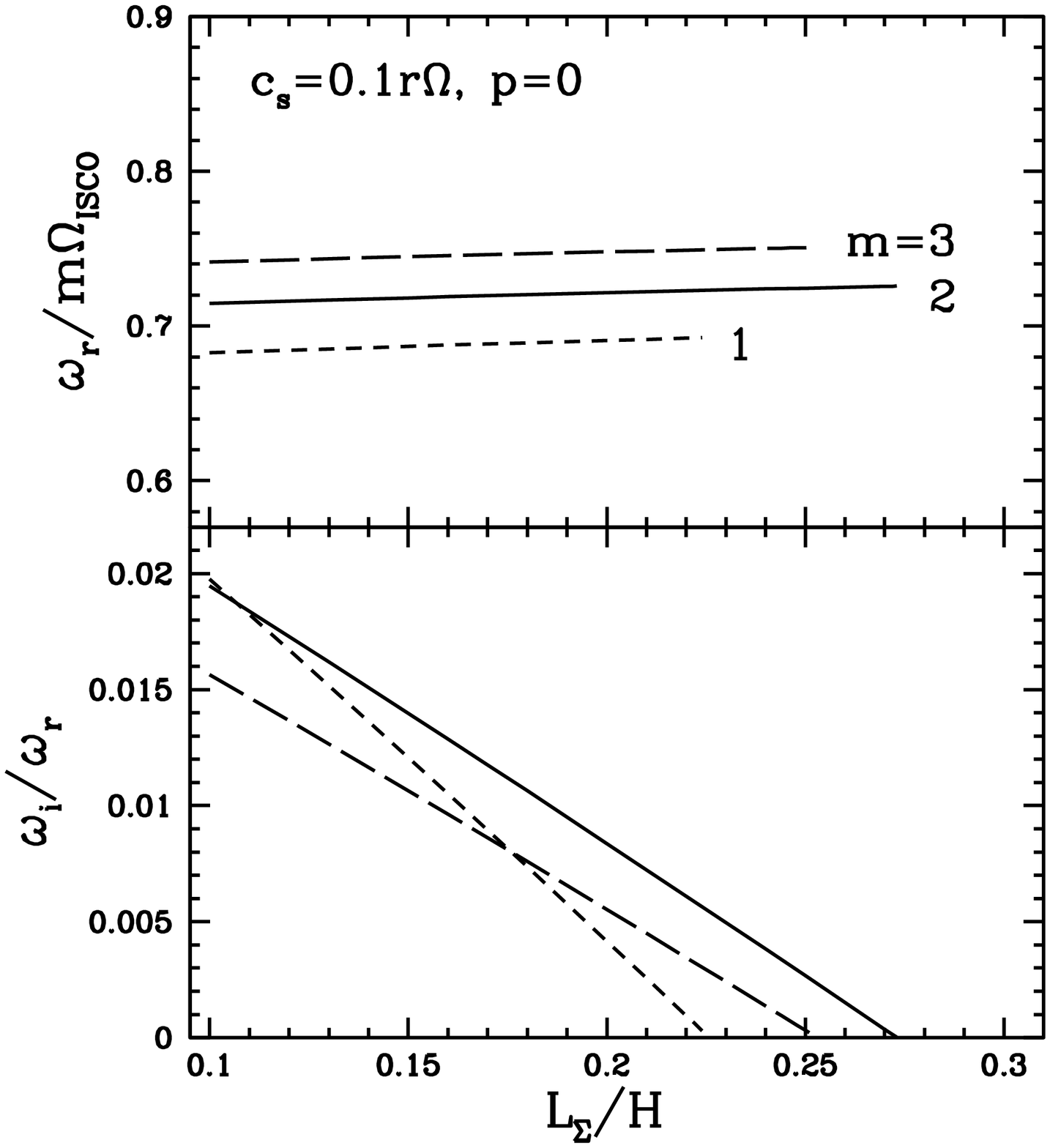,width=0.5\linewidth,clip=} &
\epsfig{file=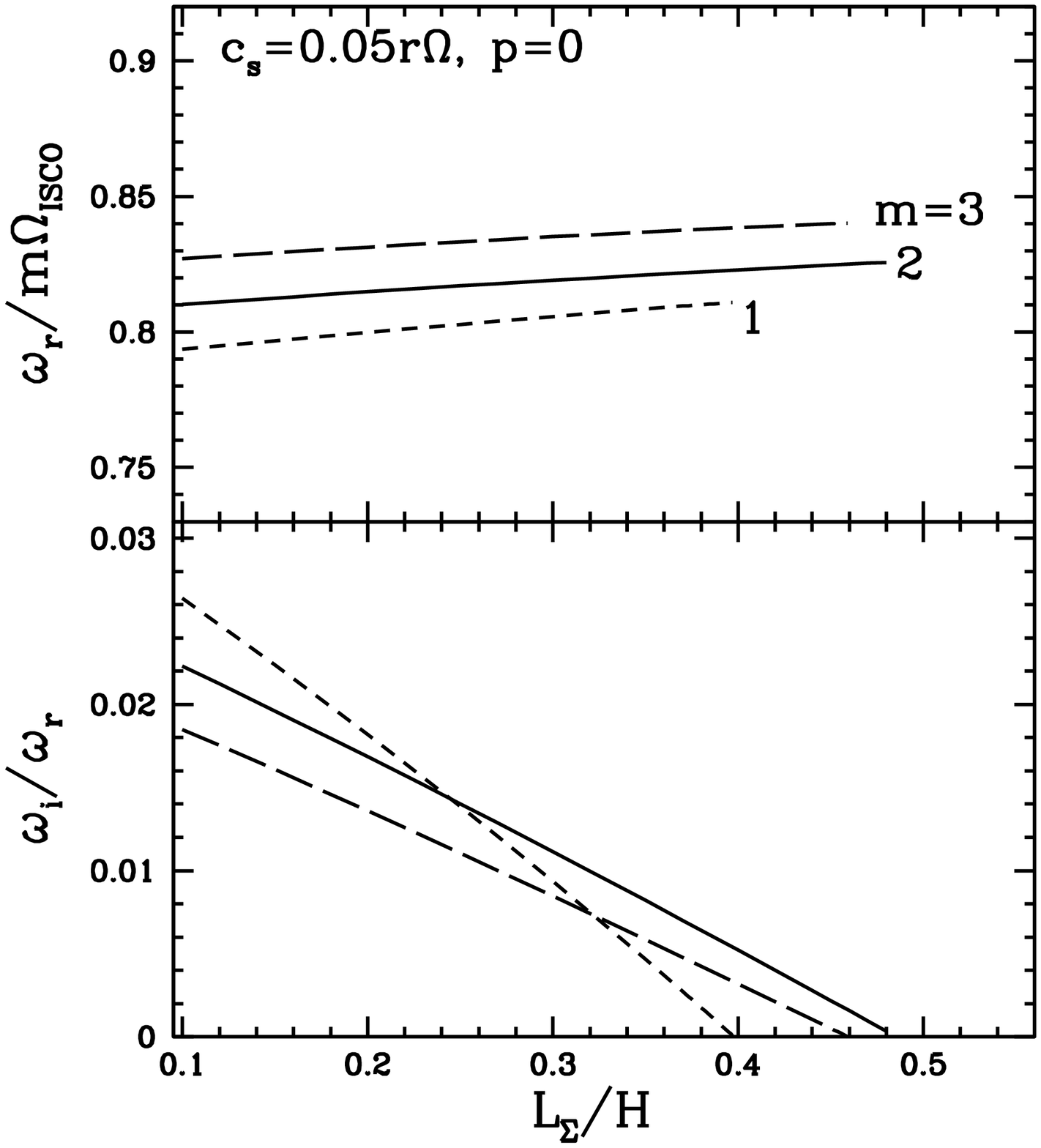,width=0.5\linewidth,clip=}
\end{tabular}
\caption{The real and imaginary frequencies of disc p-modes (with 
azimuthal wave numbers $m=1,2,3$)
as a function of $L_\Sigma/H$ [see eq.~(\ref{eq:Lc})]
The modes are calculated using the transonic inner 
boundary condition (\ref{eq:bc}) with $L_c=\infty$.
The disc has a constant surface density profile and the sound 
speed is $c_s=0.1r\Omega$ (left panels) or $0.05r\Omega$ (right panels).
}
\label{fig10}
\end{figure}

\begin{figure}
\centering
\begin{tabular}{ccc}
\epsfig{file=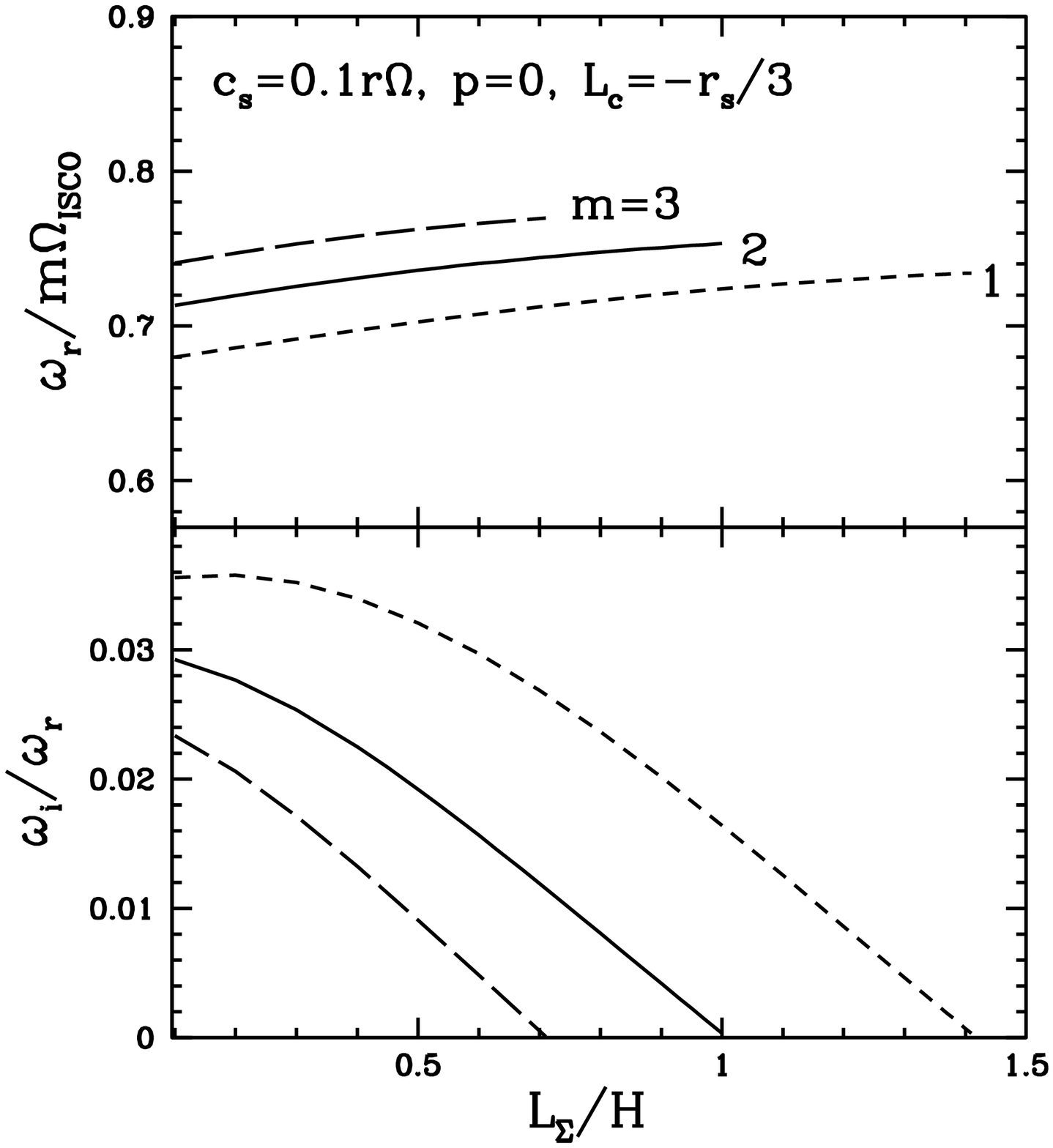,width=0.5\linewidth,clip=} &
\epsfig{file=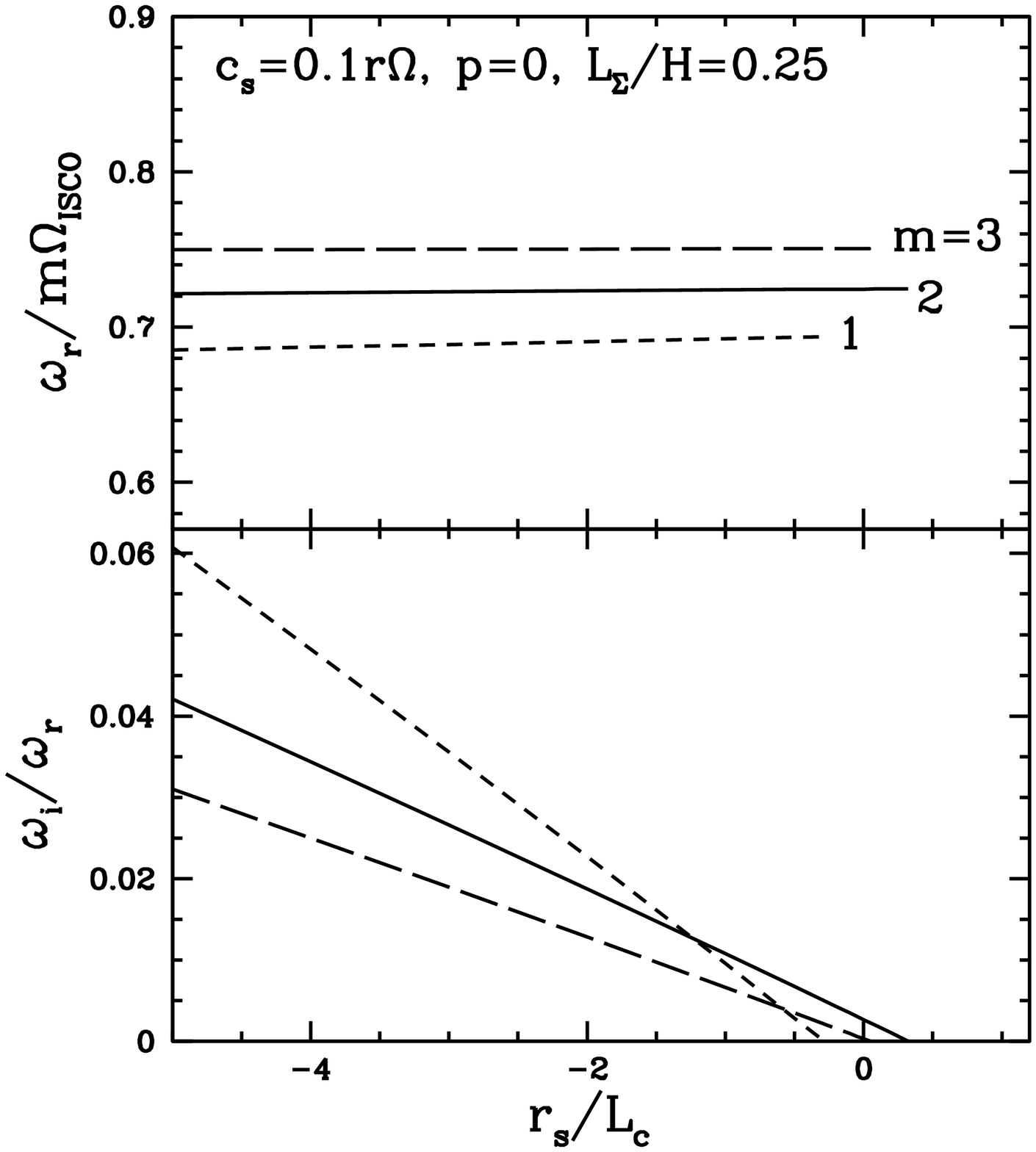,width=0.5\linewidth,clip=}
\end{tabular}
\caption{The real and imaginary frequencies of disc p-modes (with 
azimuthal wave numbers $m=1,2,3$).
The modes are calculated using the transonic inner 
boundary condition (\ref{eq:bc}). 
The disc has a constant surface density profile and the sound 
speed is $c_s=0.1r\Omega$. The left panels show the cases with
$L_c=-r_s/3$ by varying $L_\Sigma/H$, and right panels show the
cases with $L_\Sigma/H=0.25$ by varying $r_s/L_c$ 
[see eq.~(\ref{eq:Lc})].
}
\label{fig11}
\end{figure}

We solve equations (\ref{eq:sig})-(\ref{eq:uphi}) (with $u_r=0$)
subjected to the radiative outer boundary condition (\ref{eq:outer})
and the transonic inner boundary condition (\ref{eq:bc}).

Figure 9 depicts an example of the p-mode wavefunctions.
Again, the discontinuity in the angular momentum flux $F$ at 
$r_c$ signifies wave absorption; since $r_c<r_{\rm peak}$, 
this leads to mode growth.
Comparing with Fig.~4, here the angular momentum flux at $r_{\rm in}$ is
significantly nonzero, indicating wave loss through the sonic point.
Nevertheless, the corotational instability is sufficiently strong to
overcome the loss and makes the mode grow.

Figures 10--11 show the fundamental p-mode frequencies and growth rates
as a function of the disc parameters. Consistent with the result of
section 5.3 (see Fig.~8), growing modes are obtained for sufficiently
small $L_\Sigma$. Negative $L_c$ also tends to reduce wave loss
at $r_s$ and make the growing modes possible.
Such values of $L_\Sigma$ and $L_c$ are not unreasonable for 
black hole accretion discs.

It is important to note that while the mode growth rates $\omega_i$ depend
sensitively on the inner disc parameters, particularly the physical
property of the transonic flow near the ISCO, the real mode
frequencies $\omega_r$ show only weak dependence on
the inner disc parameters (e.g., $\omega_r$ decreases with increasing
sound speed; see Fig.~6).
 Thus we may expect that kHz
QPOs appear only in certain accretion states of the black hole,
and the frequencies do not vary much as the accretion rate changes.

\section{The Role of Rossby Wave Instability}

Lovelace et al.~(1999) (see also Li et al.~2000) have shown that 
when the vortensity $\zeta=\kappa^2/(2\Omega\Sigma)$ has an extremum
at a certain radius ($r_{\rm peak}$) in the disc\footnote{Lovelace et al.
considered non-barotropic flows, so the ``generalized vortensity'' depends
on the entropy profile of the disc.}, it is possible to form 
normal Rossby modes around $r_{\rm peak}$. If the trapped Rossby waves
can propagate on both sides of the corotation, a standing pattern of waves of 
opposite energies are formed, making the mode unstable --- This is the
``Rossby wave instability''. Tagger \& Varniere (2006) have considered
the MHD version of the instability and suggested that it played a role
in the diskoseismic modes around black holes (see also Tagger 2006).

We do not find any trapped Rossby modes in our calculation. To clarify the
issue in light of works by Lovelace et al. and by Tagger \& Varniere,
let us consider equation (\ref{perturbeq1}) and define the effective potential
\be
V_{\rm eff}(r)=
\frac{2m\Omega}{r\tomega}\left(\frac{d}{dr}\ln\frac{\Omega\Sigma}{D}\right)
+\frac{m^2}{r^2}+ \frac{D}{c_s^2}.
\ee
The wave equation can be approximated by $(d^2/dr^2-V_{\rm eff})\delta h\simeq 0$
(see Tsang \& Lai 2008a). We will focus on modes with $r_c$ very close to $r_{\rm peak}$ 
(i.e., $|r_c-r_{\rm peak}|\ll r_c$, so that 
$\omega_r\simeq m\Omega_{\rm peak}$; see Fig.~1). For 
$|r-r_{\rm peak}|\ll r_{\rm peak}$ in a thin disc (so that $m^2/r^2$ can be
neglected compared to $\kappa^2/c_s^2$), the effective potential becomes
\be
V_{\rm eff}(r)\simeq -\frac{2m\Omega}{r\tomega}\left(\frac{d}{dr}\ln\zeta
\right)+{\kappa^2\over c_s^2}
\simeq {2\over qL_\zeta^2}\left({r-r_{\rm peak}\over r-R_c}\right)
+{\kappa^2\over c_s^2},
\label{eq:veff2}\ee
where in the second equality we have used 
$R_c=r_c-i(r_c\omega_i/q\omega_r)$, $\Omega\propto r^{-q}$, and 
defined $L_\zeta$ via
\be
{d\ln\zeta\over dr}=-{r-r_{\rm peak}\over L_\zeta^2},\qquad ({\rm for}~
|r-r_{\rm peak}|\ll r_{\rm peak}).
\ee
Consider the case $r_c<r_{\rm peak}$ and assume $\omega_i\ll \omega_r$. 
The Rossby wave zone (where $V_{\rm eff}<0$) lies between $r_c$ and $r_c+\Delta r_R$, 
with 
\be
\Delta r_R\simeq {r_{\rm peak}-r_c\over (q/2)(L_\zeta/H)^2},
\ee
where $H\simeq c_s/\kappa$, and we have used $L_\zeta/H\sim r/H\gg 1$. The
number of wavelengths in the Rossby zone is 
\be
\int_{r_c}^{r_c+\Delta r_R}\!\!k\,dr=
\int_{r_c}^{r_c+\Delta r_R}\!\!(-V_{\rm eff})^{1/2}\,dr
\sim {4H (r_{\rm peak}-r_c)\over qL_\zeta^2}
\ee
Two points should be noted: (i) Since $\int_{r_c}^{r_c+\Delta r_R}\!k\,dr\ll 1$
for $L_\zeta\sim r_{\rm peak}$, no stationary wave can form in the Rossby zone;
(ii) since the Rossby zone lies only on one side of the corotation radius,
even if the mode can be trapped it will not grow by the Rossby wave instability
mechanism. Similar result can be obtained for the $r_c>r_{\rm peak}$ case.
We conclude that for the ``smooth'' vortensity maximum (with length scale
$L_\zeta\sim r$; see the lower panel of Fig.~1) considered in this paper,
there is no trapped Rossby mode around $r_{\rm peak}$ 
and the Rossby wave instability is ineffective.

In the hypothetical situation where the vortensity $\zeta$ has a {\it minimum}
at $r=r_{\rm min}$, equation (\ref{eq:veff2}) should be replaced by
\be
V_{\rm eff}(r)\simeq -{2\over qL_\zeta^2}\left({r-r_{\rm min}\over r-R_c}\right)
+{\kappa^2\over c_s^2},
\ee
where we have used 
\be
{d\ln\zeta\over dr}={r-r_{\rm min}\over L_\zeta^2},\qquad ({\rm for}~
|r-r_{\rm min}|\ll r_{\rm min}).
\ee
In this case, for a mode with $\omega_r=m\Omega(r_{\rm min})$
(or $r_c=r_{\rm min}$), we find 
$V_{\rm eff}(r_c)\simeq -2/(qL_\zeta^2)+1/H^2$ (for $\omega_i\ll\omega_r$). 
When $V_{\rm eff}(r_c)<0$, or when
\be
L_\zeta<\left({2\over q}\right)^{1/2}\!\!H,
\ee
Rossby waves can propagate on both sides of 
the corotation, leading to mode growth
--- this is the Rossby wave instability.
Thus, the Rossby wave instability would operate if there existed 
a ``sharp'' vortensity {\it minimum} in the disc (with $\zeta$ varying on the 
lengthscale comparable or less than the disc thickness)
--- this is not the case for typical black hole accretion discs
considered in this paper.

\section{Discussion}

High-frequency QPOs (HFQPOs) in black-hole X-ray binaries have been studied
observationally for more than a decade now and they provide a
potentially important tool for studying the strong gravitational
fields of black holes (see Remillard \& McClintock 2006).
Despite much theoretical effort,
the physical mechanisms that generate these QPOs remain unclear
(see section 1.1 for a brief review of existing theoretical models).
Ultimately, numerical simulations of realistic accretion discs
around black holes may provide the answer.
However, such simulations are still at their early stage of
development and have their own limitations (e.g., De Villiers \& Hawley 2003; 
Machida \& Matsumoto 2003, 2008; Arras et al.~2006; Fragile et al.~2007;
Reynolds \& Miller 2008; Beckwith et al.~2008; Shafee et al.~2008;
Noble et al.~2008), semi-analytical study remains a useful,
complementary approach in order to identify the key physics involved.

In this paper, we have studied the global instability of the
non-axisymmetric p-modes in black-hole accretion discs.  These modes
have frequencies $\omega\sim (0.5-0.7) m\Omega_{\rm ISCO}$ (where $m$
is the azimuthal wave number, $\Omega_{\rm ISCO}$ is the disc rotation
frequency at the inner-most stable circular orbit), where the
pre-factor (0.5-0.7) depends on the inner disc structure.  Recent
works (Arras et al.~2006; Reynolds \& Miller 2008; Fu \& Lai 2008)
suggested that, unlike other diskoseismic modes (g-modes and c-modes),
the p-modes may be robust in the presence of disc magnetic fields and
turbulence. Our linear analysis showed that due to GR effects, the
p-modes may grow in amplitude due to wave absorptions at the corotation
resonance. For a given $m$, only the lowest-order p-mode
has sufficiently high frequency ($\omega>m\Omega_{\rm peak}$; see Fig.~1) 
to be driven overstable by the corotational instability, while high-order
(lower frequency) modes are damped by the corotational wave absorption.

The greatest uncertainty of our calculation of the p-mode growth rate
concerns the boundary condition at the inner disc edge near the ISCO.
In particular, the rapid radial inflow at the ISCO has the tendency to
damp the mode (see Blaes 1986). While our analysis in section 5
indicates that this damping does not completely suppress the mode
growth under certain disc conditions, it suggests that mode growth may
not always be achieved in real black-hole accretion 
discs. Observationally, it is of interest to note that HFQPOs
are observed only when the X-ray binaries are in the steep
power-law state, while they do not appear in other spectral states
(Remillard \& McClintock 2006). In particular, HFQPOs are absent
in the thermal (soft-high) state, believed to correspond to geometrically thin 
discs extending down to the ISCO. It is reasonable to expect that 
in this state p-modes are damped due to the rapid radial inflow.

Our current understanding of the steep power-law state (also called
very high state) of black-hole X-ray binaries is rather limited.  A
thermal-radiation-emitting disc is suggested by spectral modelings,
but it is not clear whether the disc is truncated at the ISCO or
slightly larger radius (see Done et al.~2007). The observed power-law
radiation component requires a significant corona that Compton
up-scatters the disc thermal radiation. It is possible that in the
steep power-law state, the inner disc behaves as a more reflective
boundary (modeled in section 4) than a transonic flow (modeled in section 5), 
and thus more robust p-mode growth can be achieved. 
One possibility is that a significant magnetic field
flux can accumulate in the inner disc when the disc accretion rate is
sufficiently high (see Bisnovtyi-Kogan \& Lovelace 2007; Rothstein \& Lovelace 2008
and references therein). Such a magnetic field may also
enhance the corotational instability and induce variability 
in the power-law radiation flux (see Tagger \& Varniere 2006).

Although the p-mode growth rates depend sensitively on a number of
(uncertain) disc parameters (particularly those related to the inner
disc boundary), the mode frequencies are more robust (see Figs.~5-7,
10-11). More precisely, the real mode frequency can be written as
$\omega_r={\bar\omega}m\Omega_{\rm ISCO}$, where $\bar\omega<1$
depends weakly on $m$ and has only modest dependence on 
disc parameters (e.g. sound speed). This implies a
commensurate frequency ratio as observed in HFQPOs (note that in some
of our models, the $m=2,3$ modes have the largest growth rates; see
Fig.~5). The fact that $\bar\omega<1$ would also make the numerical values
of the p-mode frequencies more compatible with the measurements of the QPO
frequencies and black hole masses. 

We note that our calculations in this work are done with a pseudo-Newtonian potential. For direct comparison with observations a fully general relativistic calculation\footnote{Previous work on relativistic diskoseismic g-modes has been done by Perez et al. (1992) and Silbergleit \& Wagoner (2008) while the c-mode was studied by Silbergleit et al. (2001). Axisymmetric p-modes were studied using a general relativistic formalism by Ortega-Rodriguez et al. (2002), but these do not include the effect of the corotation singularity.} is needed including a careful treatment of the corotation singularity. Including the effect of black hole spin would likely increase the value of $\omega$ by modifying $r_{\rm ISCO}$ and $\Omega_{\rm ISCO}$, while $\bar\omega$ will likely remain similar to the non-spinning case discussed above. We plan to study these effects in future work.

\section*{Acknowledgments}

We thank Richard Lovelace for useful discussion. 
This work has been supported in part by NASA Grant NNX07AG81G, NSF
grants AST 0707628, and by {\it Chandra} grant TM6-7004X
(Smithsonian Astrophysical Observatory).



\end{document}